\documentclass[a4paper,11pt]{article}
\usepackage{jheppub}

\def\beq{\begin{equation}}
\def\eeq{\end{equation}}
\def\bea{\begin{eqnarray}}
\def\eea{\end{eqnarray}}

\def\eq#1{{Eq.~(\ref{#1})}}
\def\fig#1{{Fig.~\ref{#1}}}

\newcommand{\Lb}{\left(}
\newcommand{\Rb}{\right)}
\parskip=\itemsep               

\newcommand{\nn}{\nonumber}

\newcommand{\h}{\frac{1}{2}}

\title{Nuclei in the   toy world:\\beyond the Pomeron in zero transverse dimensions.}
\author[a,b]{Alex Kovner,}
\author[b,c]{Eugene Levin,}
\author[d]{and Michael Lublinsky}

\affiliation[a]{Physics Department, University of Connecticut, 2152 Hillside Road, Storrs, CT 06269, USA}
\affiliation[b]{Department of Particle Physics, Tel Aviv University, Tel Aviv 69978, Israel}
\affiliation[c]{Departemento de F\'isica, Universidad T\'ecnica Federico Santa Mar\'ia, and Centro Cient\'ifico-\\
Tecnol\'ogico de Valpara\'iso, Avda. Espana 1680, Casilla 110-V, Valpara\'iso, Chile}
\affiliation[d]{Physics Department, Ben-Gurion University of the Negev, Beer Sheva 84105, Israel}

\abstract{
We explore possible extensions of the $t$-channel and $s$-channel unitary model of high energy evolution in zero transverse dimensions appropriate to very high energy/atomic number where the dipole density in a toy hadron is parametrically high.
We suggest that the appropriate generalization is to allow emission of more than one dipole in a single step of energy evolution. We construct explicitly such a model that preserves the $t$-channel and s-channel unitarity and have the correct  low density limit, and study the particle multiplicity distribution resulting from this evolution. We consider initial conditions of a single dipole and many dipoles at initial rapidity. We observe that the saturation regime in this model is preceded by a parametric range of rapidities $\frac{1}{\alpha_s}\ln\frac{1}{\alpha_s}<Y<\frac{1}{\alpha_s}\ln\frac{1}{\alpha_s^2}$,  where the saturation effects are still unimportant, but multiple emissions determine the properties of the evolution. We also discuss the influence of the saturation on the
parton cascade and, in particular, find that in the saturation regime the entropy of partons becomes $S \approx \h \ln N$ where $N$ is the mean multiplicity.}

\begin{document}
\maketitle

\pagestyle{empty}

\mbox{}

\pagestyle{plain}

\setcounter{page}{1}
\author{}

\abstract{ }
\dedicated{}
\preprint{}

\section{Introduction}

The question of  asymptotic behavior of hadronic scattering amplitude at high energy has been actively studied over many decades. It is well understood by now that at high enough energy the fast growth of the scattering amplitude slows down and that the physical phenomenon that is responsible for it, is the saturation of partons in hadronic wave function. The quantitative description of this saturation however is a difficult question. We believe that the appropriate framework is the so called Reggeon Field Theory (RFT) where the effective degrees of freedom are scattering amplitude and the evolution parameter is the logarithm of energy (rapidity) \cite{BFKL,LI,glr,MUPA,MUDI,LIREV,LipatovFT,bartels,BKP,mv,MUSA,Salam,KOLE,BRN,braun,
BK}.
 Currently we know how to construct the evolution of the amplitudes and the hadronic wave function appropriate for the situation when one of the colliding hadrons may be dense, but the other one remains dilute at the energy of interest. 

Once the energy grows even further so that both hadrons have to be considered dense nontrivial modification of the current theory is necessary. This also pertains to collision of heavy ions at lower energies, where the partonic density is large already at get go. Much thought has gone into attempting to extend the current theory to this regime \cite{AKLL,KOLU1,KOLUD,SMITH,KLW,MShoshi,IAN,MUSH,LELU,KLP,LMP,LEPP}, but a consistent description of the dense regime is still wanting. 

Although a proper QCD derivation of the high energy RFT is not available, there are several constraints on the eventual form of this theory that follow from fundamental unitarity requirements. In particular, the effective theory should be $s$-channel unitary and $t$-channel unitary. 

The $t$-channel unitarity condition can be formulated in different ways. One way of putting it is the requirement that the scattering amplitude does not depend on the frame in which the scattering is described. Mathematically this is equivalent to  the property of self-duality of the RFT \cite{Kovner:2005en}. This property is built in the BFKL evolution \cite{BFKL,LI} which is appropriate for the scattering of two dilute objects. It is however lacking in the BK or JIMWLK equations \cite{BK,JIMWLK}, which describe the scattering of a dilute object on a dense one. Recently we have proposed a generalization of JIMWLK evolution which does preserve the $t$-channel unitarity \cite{KLLL1,KLLL2}.

The $s$-channel unitarity has only been discussed recently in this context. It does not have a simple mathematical formulation, but physically is equivalent to the requirement that RFT  is derivable from a fundamental unitary quantum field theory. The $s$-channel unitarity turns out to be a difficult constraint to satisfy. Neither BFKL nor JIMWLK evolution satisfy it fully, and it has not been established so far in any of the putative generalizations \cite{KLLL1}.

An interesting question to ask is whether imposing the two unitarity conditions is restrictive enough to determine the RFT completely. Or perhaps better to say, what additional guidance about RFT 
one can obtain from physical considerations beyond unitarity.

While the quantitative theory of course should be derived directly from QCD, over the years we have found simple toy models to be quite illuminating. Thus much work in early days was done on models with zero transverse dimensions in an attempt to understand general features of Pomeron interactions \cite{ACJ,AAJ,JEN,ABMC,CLR,CIAF,MUDI,RS,KLremark2,SHXI,KOLEV,BIT,nestor,LEPRI}.  Later similar zero dimensional models have been studied as a good playground to explore general features which must be present in the effective high energy theory of QCD. These universal features include $t$-channel unitarity and $s$-channel unitarity. Recently we have suggested a simple zero dimensional model which satisfies both unitarity conditions \cite{utm}.

 In this paper we further study zero dimensional case. In Section 2 we discuss the relation of the model of \cite{utm}
with a $t$-channel unitary model introduced in \cite{MUSA} and later in \cite{kl} and studied in detail in \cite{BIT}.
We show that the two models lead to identical evolution equations for the dipole probabilities, and are therefore equivalent. Below we will refer to this model UTM - unitary toy model.
We also show that the generating function frequently introduced in the framework of dipole evolution, in the RFT formulation of the same theory plays the role of the Schroedinger wave function.

In Section 3 we point out that the RFT formulation is rather flexible and allows to generalize the model in question in a way that preserves the correct unitarity properties. The new physical ingredient that enters this generalization is the observation that in one step of the evolution the emission does not have to be limited to just one gluon. In fact we argue that in the high energy dense regime one should expect emission of any number of gluons with appropriate probabilities. While in the dilute limit the emission of the additional gluons is suppressed by powers of the coupling constant, thus corresponding to NLO corrections to the emission kernel, this is not the case in the asymptotic dense regime. We construct explicitly an RFT Hamiltonian (below referred to as UTMM),  which implements such an evolution and study some of its properties. In particular we observe that it leads to a much wider distribution of dipoles in the wave function at high energy compared to the original toy model.  In both sections 2 and 3 we study the saturation effects in the parton cascades  which are not included in the JIMWLK(BK) approach. The effect of saturation are physically the same as summation of the BFKL Pomeron loops, albeit the language we use in this paper is different. This problem  has not been solved in QCD, hence the experience with the exact solvable simplified models could be useful. 

Sections 2 and 3 focus on dipole evolution of a single dipole taken as initial condition. 
In Section 4 we explore the effect of initial conditions on the probability distribution. In particular we solve for the probability distributions in UTM and UTMM 
but for $m$ dipoles in the wave function 
as initial condition. We show that, as expected, increasing $m$ shifts the asymptotic regime 
to lower rapidities. In particular if the initial number of dipoles is very large $m\sim 1/\gamma$, where $\gamma$ it the dipole-dipole scattering amplitude, the BK regime is absent in the evolution and the saturation regime dominates from the get go.

Finally  Section 5 is devoted to discussion of several  qualitative features of the models we consider.

\section{The unitary toy model (UTM)}
\subsection{The RFT formulation}
In \cite{utm} we have discussed the zero dimensional toy model defined  as an RFT.
 Mathematically the setup is the following. 
The projectile and target states of RFT are defined by the action of (projectile and target) dipole operators $d$ and $\bar d$ on the left and right vacua respectively. 
The general RFT "wave function" of the target at rapidity $Y^T$ has the form
\beq\label{psi}
|\Psi_T\rangle_{Y^T}=\sum_nP_n^T(Y^T)\bar d^n|0\rangle
\eeq
where $P_n^T(Y^T)$ are probabilities to have $n$ dipoles in the target state, $P_n^T\ge 0$, $\sum_nP_n^T=1$.
Similarly for the projectile the most general state is
\beq
\langle \Psi_P|=\sum_mP_m^P(Y^P)\langle 0|d^m
\eeq

Note that although we refer to the above objects as "wave functions", those are not wave functions of any quantum theory, but rather "wave functions" of RFT. As such their physical meaning is different from the usual  Schroedinger wave functions. In particular the coefficients in their expansion in the dipole basis are themselves physical probabilities, rather than amplitudes whose squares yield probabilities in the standard quantum mechanical setting.

In terms of these objects the scattering amplitude  is calculated as
\beq
s=\langle \Psi_P(d)|\Psi_T(\bar d)\rangle
\eeq
To calculate the overlap we use the algebra of the dipole operators
\beq \label{alg}
d\bar d=e^{-\gamma}\bar d d; 
\eeq
and the properties of the right and left "vacua"
\beq
d|0\rangle =0; \ \ \ \ \ \ \ \ \ \langle 0|\bar d=0
\eeq
The constant $e^{-\gamma}$ has the meaning of a dipole-dipole scattering matrix. In the following we assume the scattering to be weak, so that in the natural counting in powers of the coupling constant $\gamma\sim \alpha_s^2$. This is consistent with the scattering amplitude of two dipoles in QCD. With the assumption that $\gamma$ is small, and will freely use $e^{-\gamma}\approx 1-\gamma$ whenever convenient.

Note that the algebra of $d$ and $\bar d$ can be represented explicitly on functions of $\bar d$ 
by \footnote{we have used the symbol of partial derivative in order to avoid confusion between the differential $d$ and the dipole operator $d$.}
\beq \label{repr} d=\exp\{-\gamma \bar d\frac{\partial}{\partial\bar d}\}
\eeq
As a consequence
\beq
\langle 0|d^m\bar d^n|0\rangle=e^{-\gamma mn}
\eeq
and
\beq \label{S}
s(Y)=\sum_{m,n}e^{-\gamma mn}P_m^P(Y_0)P_n^T(Y-Y_0)\,\,\equiv \,\,\sum_{m,n}\,\sigma^{m n}P_m^P(Y_0)P_n^T(Y-Y_0)\eeq
with $\sigma = e^{- \gamma}$.
In this expression the total rapidity (logarithm of energy) of the scattering process is $Y$, while $Y_0$ defines the frame in which the calculation (observation) is performed.
The physical amplitude of course should be independent of $Y_0$ and depend  on $Y$ only.

The energy evolution of the scattering amplitude is given by the evolution equation generated by an RFT Hamiltonian according to
\beq 
s(Y)\,=\,\langle \Psi_P(d)|e^{-HY}|\Psi_T(\bar d)\rangle
\eeq
where $H$ is an operator function of $d$ and $\bar d$.

The unitarized toy model (UTM) of \cite{utm} is defined by the Hamiltonian

\beq \label{UTM}
H_{UTM}=-\frac{\Delta}{\gamma} \bar P P
\eeq
where the Pomeron operators are related to dipoles as
\beq P=1-d;\ \ \ \ \ \bar P=1-\bar d.
\eeq

The Hamiltonian generates the evolution of the RFT wave function. For a $n$ dipole target state evolved by an infinitesimal rapidity $\delta Y$
\beq \label{evolut}
e^{-H_{UTM}\delta Y}|n)\rangle\approx \left(1-\delta Y\frac{\Delta}{\gamma}\left[1-e^{-\gamma n}\right]\right)|n\rangle+\delta Y\frac{\Delta}{\gamma}\left[1-e^{-\gamma n}\right]|n+1\rangle
\eeq
The constant $\Delta$ which determines the probability to emit a dipole in one step of the evolution is a model parameter. In terms of the counting of powers of the coupling constant we will set it to be of order $\Delta\sim \alpha_s$, which is consistent with QCD. Thus our two model parameters are both small, and $\gamma\sim \Delta^2$.

The evolution \eq{evolut} has several important properties. First, it is $s$-channel unitary. This is obvious since a step in the evolution generates a new dipole state with positive probability. It is also $t$-channel unitary, as can be seen by explicit derivation of the evolution of a target state, which turns out to be identical to eq.(\ref{evolut}), \cite{utm}. The $t$-channel unitarity is assured by the self duality of the Hamiltonian. i.e. invariance under the transformation  $d\rightarrow \bar d$ accompanied by the exchange of order of the factors $d$ and $\bar d$, which we will refer to as transposition.
Explicitly, the duality transformation  is
\beq  
H(d,\bar d)\rightarrow H^T(\bar d, d)
\eeq
Finally, another important point is that the probability of emission of an extra dipole for large $n$ does not depend on $n$, since $1-e^{-\gamma n}\rightarrow 1$ for $n\gg 1/\gamma$. Thus although for small $n$ (small rapidity) the number of dipoles grows exponentially, at large $n$ the growth is much slower. This feature of saturation is what we expect from the saturation in QCD as well. At large rapidity the cross section is dominated by configurations with large $n$. Thus the probability to emit an extra dipole is constant and the evolution becomes similar to a random walk in the dipole number space. In this sense the evolution saturates at high energy. We will observe this property explicitly below.

The quantum evolution can be cast in the form of the Schroedinger equation of the RFT. Let us define the target RFT wave function in the $\bar d$ representation(an identical discussion holds for the projectile), i.e.
\beq\label{z}
|\Psi\rangle=Z(\bar d)|0\rangle
\eeq
The proper normalization of the RFT wave function is not given by the usual integral condition as for a Schroedinger wave function, but rather by 
\beq \label{norm} Z(1)=1 \, \eeq
The Schroedinger equation for $Z$ is derived by acting with the Hamiltonian eq.(\ref{UTM}) while utilizing the algebra
eq.(\ref{alg}) and eq.(\ref{repr}):
\beq\label{sch}
\frac{\partial}{\partial Y}Z^{\mbox{\tiny  UTM}}_Y(u)=-\frac{\Delta}{\gamma}(1-u)\left(1-e^{-\gamma u\frac{\partial}{\partial u}}\right)Z^{\mbox{\tiny  UTM}}_Y(u)=\,\frac{\Delta}{\gamma}( u - 1) \Bigg( Z^{\mbox{\tiny  UTM}}_Y(u) - \,Z^{\mbox{\tiny  UTM}}_Y\Lb e^{-\gamma}\, u\Rb\Bigg)
\eeq

Note that $Z$ is precisely the probability generating function as it is frequently defined in the framework of similar reaction-diffusion models.
The standard definition of the generating function is
\beq \label{Zi}
Z_Y(u)\equiv \sum_nP_n(Y)u^n
\eeq
so that
\beq P_n(Y)=\frac{1}{n!}\frac{\partial^n}{\partial u^n}Z(Y)|_{u=0}
\eeq
Comparing this definition with eq.(\ref{z}) and using eq.(\ref{psi}) we see that the  two functions are indeed identical.

We note that sometimes  rather than calculating probabilities $P^{\mbox{\tiny  UTM}}_n$ it is more useful to calculate factorial moments of the probability distribution defined as
\beq
M_k\equiv \langle n(n-1)...(n-k+1)\rangle\equiv\sum_{n=0}^{\infty}n(n-1)...(n-k+1)P_n\,=\,\sum_{n=0}^{\infty}{n!\over (n-k)!}\,P_n\eeq
 These moments can be calculated from the generating function $Z$ as 
\beq
M_k=\frac{\partial^k}{\partial u^k}Z(u)|_{u=1}
\eeq
which is equivalent to the representation
\beq
Z_Y(u)=1+\sum_{k=1}^\infty\frac{1}{k!}M_k(Y)(u-1)^k
\eeq
 \subsection{The frame invariant formulation}
An alternative approach to defining dipole models of this type was discussed a while ago in \cite{MUSA}, and later in \cite{kl} and \cite{BIT}. The starting point of these works is  explicit invariance of the evolution equation for probabilities $P_n$ under the change of Lorentz frame. 

One starts with the eq.(\ref{S}) and requires that the evolution of the probabilities is such that the expression for the s-matrix does not depend on the frame in which it is calculated, i.e. on the value of $Y_0$. If in addition one assumes that only one dipole is emitted in one step of the evolution, i.e. 
\beq
\frac{d}{dY}P_n(Y)=f_nP_n(Y)+g_nP_{n-1}(Y)
\eeq
one finds that the only solution compatible with the dilute limit is
\beq \label{1}
\frac{d P^{\mbox{\tiny  UTM}}_n(Y)}{ d Y}\,=\,- \frac{\Delta}{\gamma} \Lb 1\,-\,e^{- \gamma n}\Rb P^{\mbox{\tiny  UTM}}_n(Y) \,\,+\,\,
\frac{\Delta}{\gamma} \Lb 1\,-\,e^{- \gamma(n - 1) }\Rb\,P^{\mbox{\tiny  UTM}}_{n-1}(Y)
\eeq


Both models, \eq{UTM} and \eq{1}  describe the evolution of the same system - an ensemble of dipoles (colorless "partons"). In fact, although it was not realized in \cite{utm}, the two are equivalent. To see this we should recast the evolution of the RFT wave function in terms of the evolution of probabilities. Starting with eq.(\ref{evolut}) and reinterpreting it as evolution of probabilities we indeed immediately obtain eq.(\ref{1}).
Another way to see this equivalence is to start with the probability evolution eq.(\ref{1}) and derive the evolution of the generating function defined as eq.(\ref{Zi}). This equation is identical  with eq.(\ref{sch}) which again demonstrates that the two models are identical.

It is interesting to understand the main properties of the probability distribution as it evolves from lower to higher rapidities. We will do that in the rest of this section. We note that some of these results have already appeared before, e.g. in \cite{BIT}. We present them here for completeness, as well as to set up the stage for generalizing the model in the next section.

\subsection{The BFKL-BK limit}

When one of the colliding objects is dilute, the model above reduces to the zero dimensional BK model. For dilute target, for example the scattering amplitude of each projectile dipole is small, and one can formally expand $d$ in power series in $\gamma$
\beq
d\approx 1-\gamma \bar d\frac{\partial}{\partial\bar d}
\eeq
The Hamiltonian then becomes
\beq
H_{BK}=\Delta\left[\bar d^2-\bar d\right]\frac{\partial}{\partial \bar d}
\eeq
which is precisely the BK Hamiltonian.

Analysis of this limit is quite straightforward if we restrict ourselves to the properties of the dilute object. As was explained in \cite{utm}, the BK Hamiltonian violates $s$-channel unitarity if applied to the wave function of the dense projectile, and  we are not going to consider this case. On the other hand when evolving the state of the dilute target the BK evolution of the wave function is equivalent to the BFKL cascade. With this in mind we will allow ourselves to refer to the resulting probability distribution interchangeably as either BK or BFKL.

 First off, it is easy to see that the Schoedinger equation becomes
\beq\label{ZBK}
\frac{\partial}{\partial Y}Z^{BK}_Y(u)\,\,=\,\,- \Delta\, u\,(1-u)\, \frac{\partial}{\partial\, u} Z^{BK}_Y(u)
\eeq
This is easily solved noting that it can be rewritten in a simple way as 
\beq
\left[\frac{\partial}{\partial Y}-\frac{\partial}{\partial t}\right]Z^{BK}=0
\eeq
where
\beq
t=\frac{1}{\Delta}\ln\frac{1-u}{u}
\eeq
Thus for any initial condition $Z^{BK}_0(u)$ the solution is
\beq\label{zsol}
Z^{BK}_Y(u)=Z^{BK}_0\left(\frac{u}{u(1-e^{\Delta Y})+e^{\Delta Y}}\right)
\eeq

The most common case considered in the literature is when at initial energy the target contains one single dipole, $Z^{BK}_0=u$. In this and the next section we will concentrate on solutions that correspond to this initial condition. We will consider the case of multiple dipoles at initial rapidity in Section 4.  

For a single dipole initial condition  the solution at rapidity $Y$ is
\beq
Z^{BK}_Y(u;1)=\frac{u}{u(1-e^{\Delta Y})+e^{\Delta Y}}
\eeq
Using this generating function it is easy to see that (here the index $_{(1)}$ indicates the number of dipoles at initial rapidity) 
\beq\label{m1}
M^{BK}_{1(1)}(Y)\equiv N(Y)=e^{\Delta Y}
\eeq
and 
\beq \label{PNBK}
P_{n(1)}^{BK}(Y)=\frac{1}{N(Y)-1}\left(1-\frac{1}{N(Y)}\right)^{n}
\eeq
The higher factorial moments for this distribution have a simple structure:
\beq \label{MKBK}
M^{BK}_{k(1)}\Lb Y\Rb\,\,\,=\,\,k! \,N(Y) \Lb N(Y)\,\, - \,\,1\Rb^{k-1}
\eeq
At high energy where $e^{\Delta Y}\gg 1$ these become
\beq\label{PNBK1}
P_{n(1)}^{BK}(Y)\rightarrow \frac{1}{N(Y)}e^{-\frac{n}{N(Y)}};
 \ \ \ \ \ \ M^{BK}_{k(1)}(Y)\rightarrow k!N^k(Y)
 \eeq

\subsection{The UTM probability distribution}
We now  turn to the probability distribution in UTM beyond the BFKL-BK limit.
Various properties of UTM were studied in depth in \cite{BIT}. We mention that the equation for probabilities can be explicitly solved. In particular for the initial condition of a single dipole $P_1(0)=1; \ \ P_{n\ne 1}(0)=0$ the solution is
\begin{equation}\label{PN}
P^{\mbox{\tiny  UTM}}_1(Y)=e^{-\omega_1Y}; \ \ \ \ \ \ P^{\mbox{\tiny  UTM}}_{n>1}(Y)= \oint \frac{d \omega}{2 \pi i}\,e^{\omega\,Y}\,\, \frac{1}{\omega_n} \prod^n_{k=1} \frac{\omega_k}{\omega\,+\,\omega_k}\,\,=\,\,
\prod_{j=1}^{n-1}\omega_j\sum_{i=1}^{n}\prod_{k\ne i, k=1}^{n}{\frac{1}{\omega_k-\omega_i}e^{-\omega_iY}}
\end{equation}
with
 \begin{equation}\label{OMN}
 \omega_n= \frac{\Delta}{\gamma}[1-e^{-\gamma n}]
 \end{equation}
 Although these formulae are explicit, they do not give one directly an understanding of the properties of the distribution. To get a better idea about the importance of the saturation corrections we first consider the limit of very large energy.
 \subsubsection{The asymptotic distribution at $Y\rightarrow\infty$}
 
To find the behavior of the probabilities at high energy  we note that 
for large $Y$ we expect $Z_Y\Lb e^{-\gamma} u\Rb \,\ll\,Z_Y\Lb u\Rb$, since on average the number of dipoles is expected to be large, which means that high powers of $u$ are most important in the generating function. Thus to find the large $Y$ asymptotics we can drop the second term in the Schroedinger equation eq.(\ref{sch}). The resulting equation is simple 
\beq\label{schsat}
\frac{\partial}{\partial Y}Z^{\mbox{\tiny  UTM}}_Y(u)=\,\frac{\Delta}{\gamma}( u - 1)  Z^{\mbox{\tiny  UTM}}_Y(u) \eeq
and yields  the asymptotic solution 
\beq \label{SOL0}
Z_Y^{asymp}\Lb u\Rb\,\,=\,\,e^{  \frac{\Delta}{\gamma}( u - 1) \,(Y-Y_0)}Z_{Y_0}(u)
\eeq
Here $Y_0$ is the rapidity starting from which we can use the asymptotic equation, and
the function $Z_{Y_0}(u)$ is determined by the initial condition and the evolution up to the rapidity $Y_0$. For very large rapidity $Y\gg Y_0$ the exact value of $Y_0$ does not matter. Likewise the initial condition should not significantly affect the properties of the distribution. Formally we assume that the function $Z_{Y_0}(u)$ describes states with relatively small number of particles. It therefore is dominated by small powers of $u$ and is a relatively smooth function which can be approximated by a constant. The normalization \eq{norm} then sets this constant to unity. Thus the asymptotic generating function can be approximated by
\beq \label{SOL11}
Z_Y^{asymp\,\mbox{\tiny  UTM}}\Lb u\Rb\,\,\approx\,\,e^{  \frac{\Delta}{\gamma}( u - 1) \,Y}
\eeq
Calculating the probabilities we find the Poisson distribution :
\beq \label{PD}
P^{asymp\,\mbox{\tiny  UTM}}_n\,=\,P^{PD}_n\Lb N\Rb\,\,=\,\,\frac{N^n(Y)}{n!}e^{-N(Y)}
\eeq
with the average multiplicity 
\beq N(Y) = \frac{\Delta}{\gamma}\,Y
\eeq  
This distribution can also be directly  obtained from the general solution of \eq{PN} replacing  $\omega_n \,\rightarrow\,\frac{\Delta}{\gamma}$ which is appropriate for $n$ large such that $n\gamma\gg 1$.

The factorial moments for the Poisson distribution are
\beq 
M^{PD}_k(Y)=N^k(Y)
\eeq

We note that the properties of this probability distribution are significantly different from that in the BFKL cascade.
The average number of dipole  grows only linearly with rapidity rather than exponentially. This is a direct consequence of saturation of the emission amplitudes in eq.(\ref{evolut}) at large $n$. As noted above, the probability for emission of an extra dipole at large rapidity is a constant and does not depend on the number of dipoles already present in the wave function. As a result the evolution is similar to random walk and the average number of dipoles grows only linearly in rapidity. For BK evolution, where the emission probability is proportional to the number of dipoles present, the growth is exponential as reflected in eq.(\ref{m1}).
 Another important difference is that 
the distribution \eq{PD} unlike \eq{PNBK1} does not obey  KNO scaling \cite{KNO,KNO1,KNO2} and decreases much faster at large values of 
$n>N(Y)$ compared to $P^{BK}$.


   \subsubsection{The UTM  parton cascade at "intermediate" $n$. }
   The Poisson distribution derived above is valid at asymptotically large energy. What about  the intermediate region, where the number of particles is large enough so that the BK limit is not valid, but the energy is still not asymptotically large?  We probe this regime assuming that $P_n^{\mbox{\tiny  UTM}}$ is a smooth function of $n$, i.e. we replace 
   $P^{\mbox{\tiny  UTM}} _{n-1}(Y) $  in \eq{1} by
   \beq \label{PL1}
 P^{\mbox{\tiny  UTM}}_{n-1}(Y)  \,\,=\,\,P^{\mbox{\tiny  UTM}}_n(Y) \,-\,\frac{ \partial P^{\mbox{\tiny  UTM}}_n(Y)}{\partial n}
 \eeq
  which assumes $\frac{ \partial^2 P^{\mbox{\tiny  UTM}}_n(Y)}{\partial n^2}   \,\,\ll\,\,\frac{ \partial P^{\mbox{\tiny  UTM}}_n(Y)}{\partial n} $. 
  This assumption will have to be checked {\it a posteriori} given the solution.
In this approximation   \eq{1} takes the form
   \beq \label{PL2}
   \frac{\partial P^{\mbox{\tiny  UTM}}_n(Y)}{\partial Y}\,\,=\,\,-\frac{\Delta}{\gamma}\Lb \Lb 1 - e^{- \gamma}\Rb\,e^{ - (n - 1) \gamma} \,P^{\mbox{\tiny  UTM}}_n(Y) 
   \,\,+\,\,\Lb 1 - e^{- (n - 1) \gamma}\Rb \,  \frac{ \partial P^{\mbox{\tiny  UTM}}_n(Y)}{\partial n}\Rb
   \eeq
   
\begin{figure}[ht]
    \centering
  \leavevmode
  \begin{tabular}{c c}
      \includegraphics[width=8.5cm]{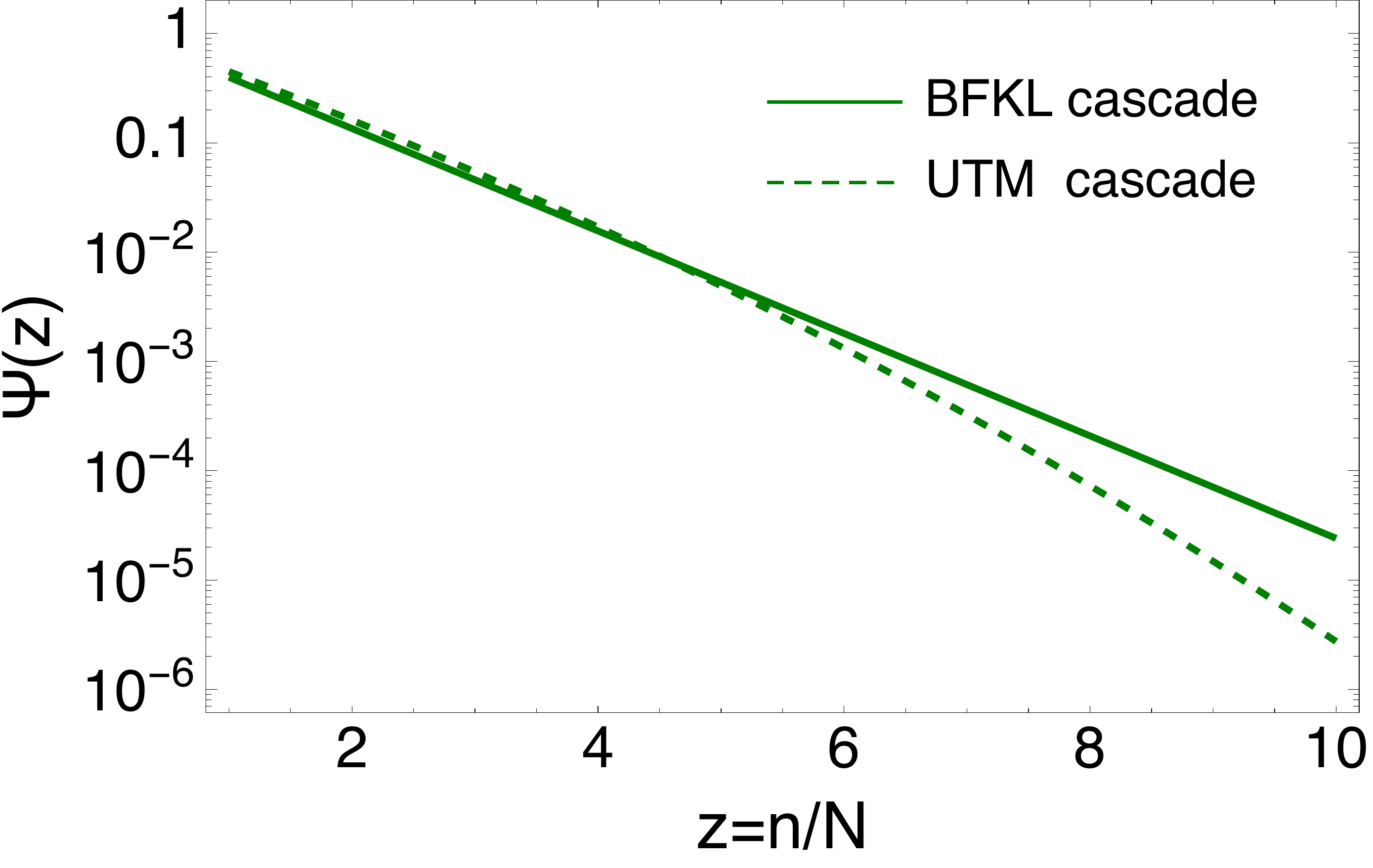}  &  \includegraphics[width=8.5cm]{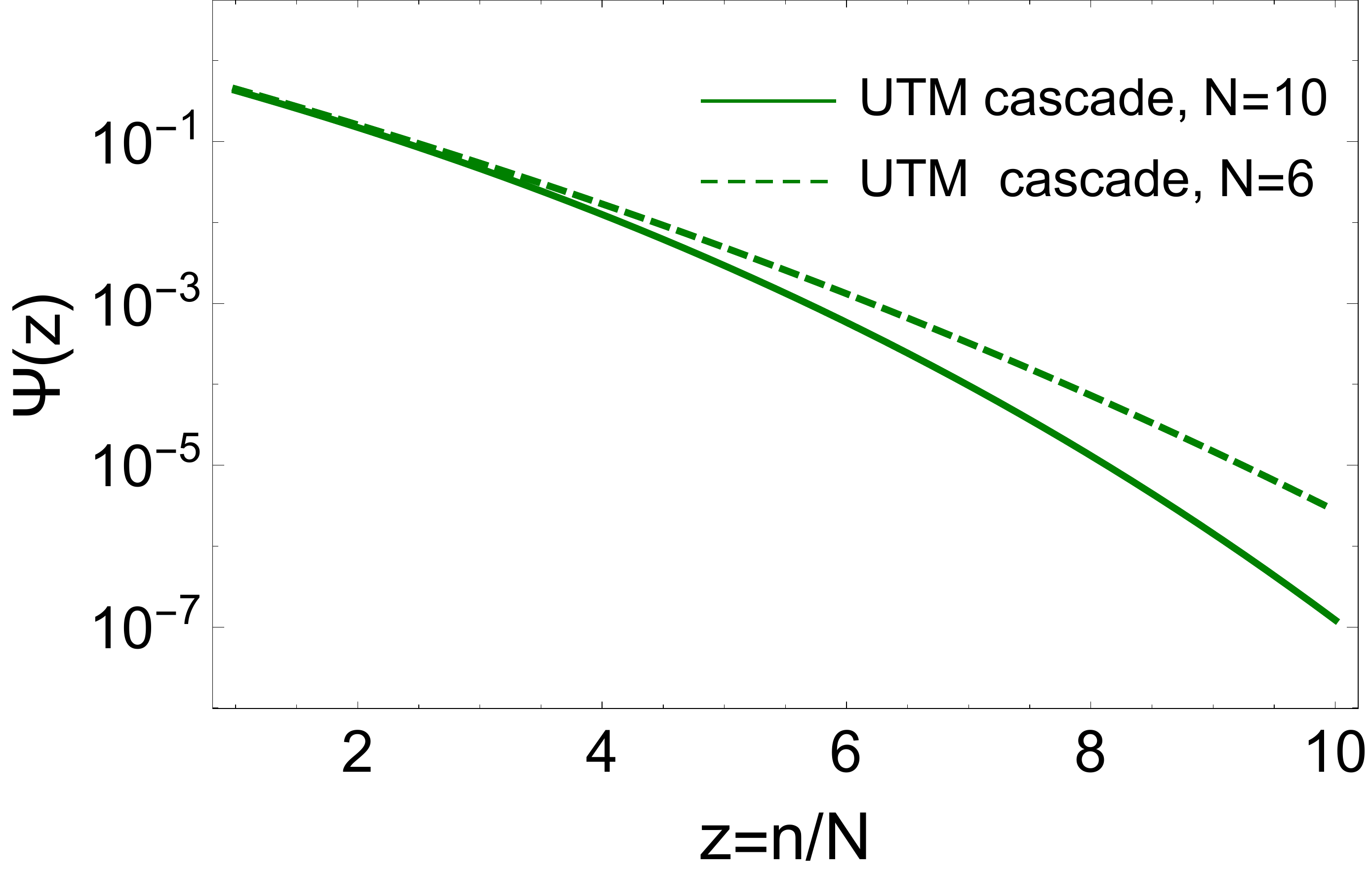}\\
      \fig{comp}-a &\fig{comp}-b\\
      \end{tabular}
      \caption{\fig{comp}-a: The KNO function  $\Psi\Lb z = \frac{n}{N}\Rb\,\,=\,\,N P_n$ for 
   the BFKL cascade (\eq{PNBK}) and the UTM cascade (\eq{PL100a}). Here $N$ is the average multiplicity for the corresponding distribution, which is taken as N=6 in the figure. \fig{comp}-b shows that the KNO scaling for UTM is only approximate and  the actual multiplicity distributions has more complex dependence on the mean multiplicity. }
\label{comp}
   \end{figure}

   We write the solution in the form $P^{\rm UTM}_n\Lb Y\Rb \,\,=\,\,\hat{P}_n\,\widetilde{P}_n\Lb Y\Rb  $, where $\hat{P}_n$
   is a particular solution of the equation:
     \beq \label{PL3}
   \Lb 1 - e^{- \gamma}\Rb\,e^{ - (n - 1) \gamma} \,\hat{P}_n 
   \,\,+\,\,\Lb 1 - e^{- (n - 1) \gamma}\Rb \,  \frac{ \partial\hat{ P}_n}{\partial n} \,\,=\,\,0.
   \eeq  
   This is solved by\footnote{We approximate $\exp\{-\gamma\}\approx 1-\gamma$.}
      \beq \label{PL4}
 \hat{P}_n\,\,=\,\, \exp\Lb- \frac{1  - \exp\Lb - \gamma\Rb}{\gamma}  \ln  \Lb 1 - e^{- (n - 1) \gamma}\Rb\Rb \, \,\,\approx\,\,\frac{1}{\Lb 1 - e^{- (n - 1) \gamma}\Rb}.\eeq     
  The equation for $\widetilde{P}_n(Y)$  then becomes:
      \beq \label{PL5}
   \frac{\partial \widetilde{P}_n(Y)}{\partial Y}\,\,=\,\,-\frac{\Delta}{\gamma}\,\Lb 1 - e^{- (n - 1) \gamma}\Rb \,  \frac{ \partial \widetilde{P}_n(Y)}{\partial n}.
   \eeq   
  The general solution of \eq{PL5}  is an arbitrary function of $\Lb \Delta Y  \,+ \,f(n)\Rb$ with $f(n)$ satisfying:
  \beq \label{PL6}
  \frac{d \,f(n)}{d n} \,\,=\,\,-\,\frac{
 \gamma}{ 1 - e^{-(n - 1)\,\gamma}}
 \eeq
 or
 \beq \label{PL7}
 f(n) \,\,=\,\, -\,\ln\Lb e^{(n - 1)\,\gamma} \,\,-\,\,1\Rb
 \eeq
  Hence 
  \beq \label{PL8}
  P^{\mbox{\tiny  UTM}}_n\Lb Y\Rb\,\,=\,\, \frac{1}{ \Lb 1 - e^{- (n - 1) \gamma}\Rb} F\Bigg( \zeta(Y,n)\Bigg)
  \eeq
  where we have defined 
  \beq
    \zeta(Y,n)\,\,=\,\,- \Delta\,Y \,+\,\ln\Bigg( \frac{ e^{(n - 1) \gamma}\,\,-\,\,1}{\gamma} \Bigg).\label{PL101}
    \eeq
 To find the function $F$ that corresponds to a particular initial condition we need to match it to the solution of BK equation at small $n\gamma$ calculated with the same initial condition. For the evolved single dipole 
matching with  the solution of \eq{PNBK1} we obtain
   \begin{subequations}    
  \bea
  P^{\mbox{\tiny  UTM}}_{n(1)}\Lb Y\Rb\,\,&=&\,\,\frac{\gamma}{\Lb 1 - e^{- (n - 1) \gamma}\Rb}\,\exp\Bigg( - e^{\zeta(Y,n)} \,\,+\,\,\zeta(Y,n)\Bigg) \label{PL100a}\\
  \,\,&\xrightarrow{n\gamma<1}&\,\,e^{-\Delta Y}e^{-ne^{-\Delta Y}}
     \eea
  \end{subequations}
 In \fig{comp} we compare  $P_n$ of the BFKL cascade (\eq{PNBK}) and of the UTM cascade with saturation given by \eq{PL100a}.
  
To determine the range of validity of this calculation we consider
\beq\label{p2p}
\frac{\partial P^{\mbox{\tiny  UTM}}_{n(1)}}{\partial n}=\left[\gamma-e^{-\Delta Y+(n-1)\gamma}\right]P^{\mbox{\tiny  UTM}}_{n(1)}; \ \  \frac{\partial^2P^{\mbox{\tiny  UTM}}_{n(1)}}{\partial n^2}=\left[-\gamma e^{-\Delta Y+(n-1)\gamma}+\left[\gamma-e^{-\Delta Y+(n-1)\gamma}\right]^2\right]P^{\mbox{\tiny  UTM}}_{n(1)}
\eeq
First, we note that that at high enough rapidity, $Y>\frac{1}{\Delta}\ln \frac{1}{\gamma}$ the probability distribution has a maximum at
\beq\label{nmax1}
n_{max}-1=\frac{\Delta}{\gamma}Y-\frac{1}{\gamma}\ln\frac{1}{\gamma}
\eeq
Second, we see that for any fixed $n$ at large enough $Y$ we have $\frac{d^2P_n}{dn^2}\ll \frac{dP_n}{dn}$ and therefore our approximation is valid. Moreover in the vicinity of the maximum $n\sim n_{max}$ the calculation is valid at any rapidity. As is clear from eq.(\ref{p2p}), parametrically the range of validity of the present approximation is given by
\beq\label{limit} n-1<\frac{\Delta}{\gamma}Y
\eeq
One can write down an approximate expression for the average multiplicity in the distribution $P_{n(1)}^{\mbox{\tiny  UTM}}$. Approximating the sum over $n$ by the integral while calculating $N(Y)$ we find
\beq \label{PL11}
N(Y)\equiv M_1(Y)\,\,\approx\,\,\frac{1}{\gamma} e^{\frac{1}{\gamma}e^{-\Delta Y}} \Gamma\Lb 0, \frac{1}{\gamma}e^{-\Delta Y}\Rb
 \eeq
where $\Gamma(x,z)$ is an (upper) incomplete $\Gamma$-function\cite{RY}. 

 From \eq{PL11} one can see that  $N(Y)$  reduces to the BFKL cascade value  $N(Y)\,\to\,\exp\Lb \Delta \,Y\Rb$  for small $Y$. On the other hand for large $Y$  such that $e^{-\Delta Y}\ll \gamma$  \eq{PL11} yields  $N(Y)\,\,\to\,\frac{\Delta}{\gamma} \,Y-\frac{1}{\gamma}\ln\frac{1}{\gamma}$  consistent with the value of $n_{max}$ found in eq.(\ref{nmax1}). Interestingly this result for the average multiplicity is also consistent with the asymptotic limit of \eq{PD}, even though for very large $Y$ our formal discussion above indicates that the range of validity of the approximation eq.(\ref{PL1}) is limited by   eq.(\ref{limit}).

\begin{figure}
    \centering
  \leavevmode
      \includegraphics[width=12cm]{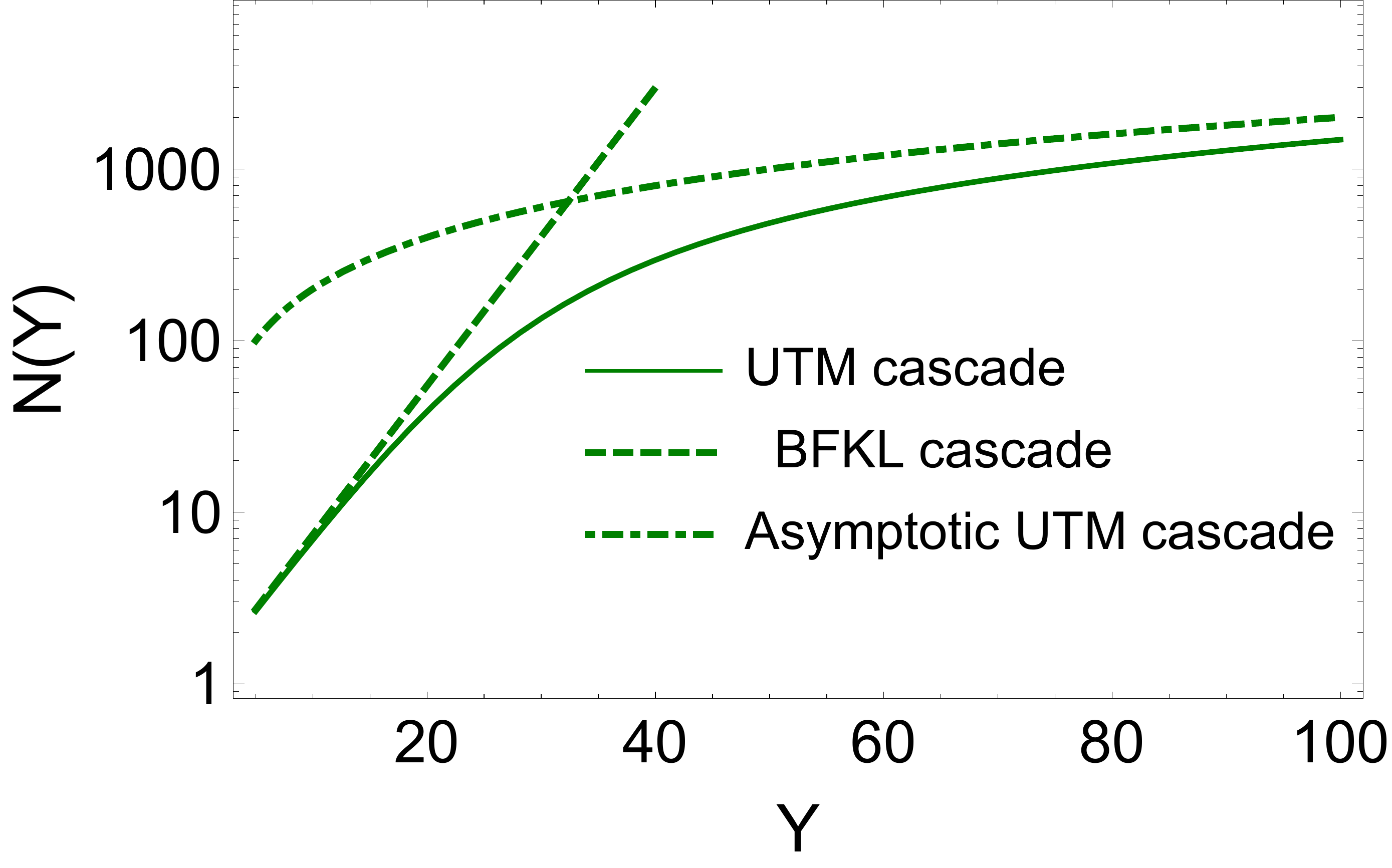}  
      \caption{The mean multiplicity $N(Y)$ (see \eq{PL11})  versus $Y$ (solid curve). 
         The mean multiplicity of the BFKL cascade is equal to $ \exp\Lb \Delta\,Y\Rb$, the asymptotic multiplicity for the UTM cascade (see \eq{PD}) is taken as $N^{asymp}(Y) = \frac{\Delta}{\gamma} \,Y$.  $\Delta = 0.2$ and $\gamma = 0.01$. }
\label{utmmult}
   \end{figure}

\begin{figure}[ht]
    \centering
  \leavevmode
  \begin{tabular}{c c}
      \includegraphics[width=8.5cm]{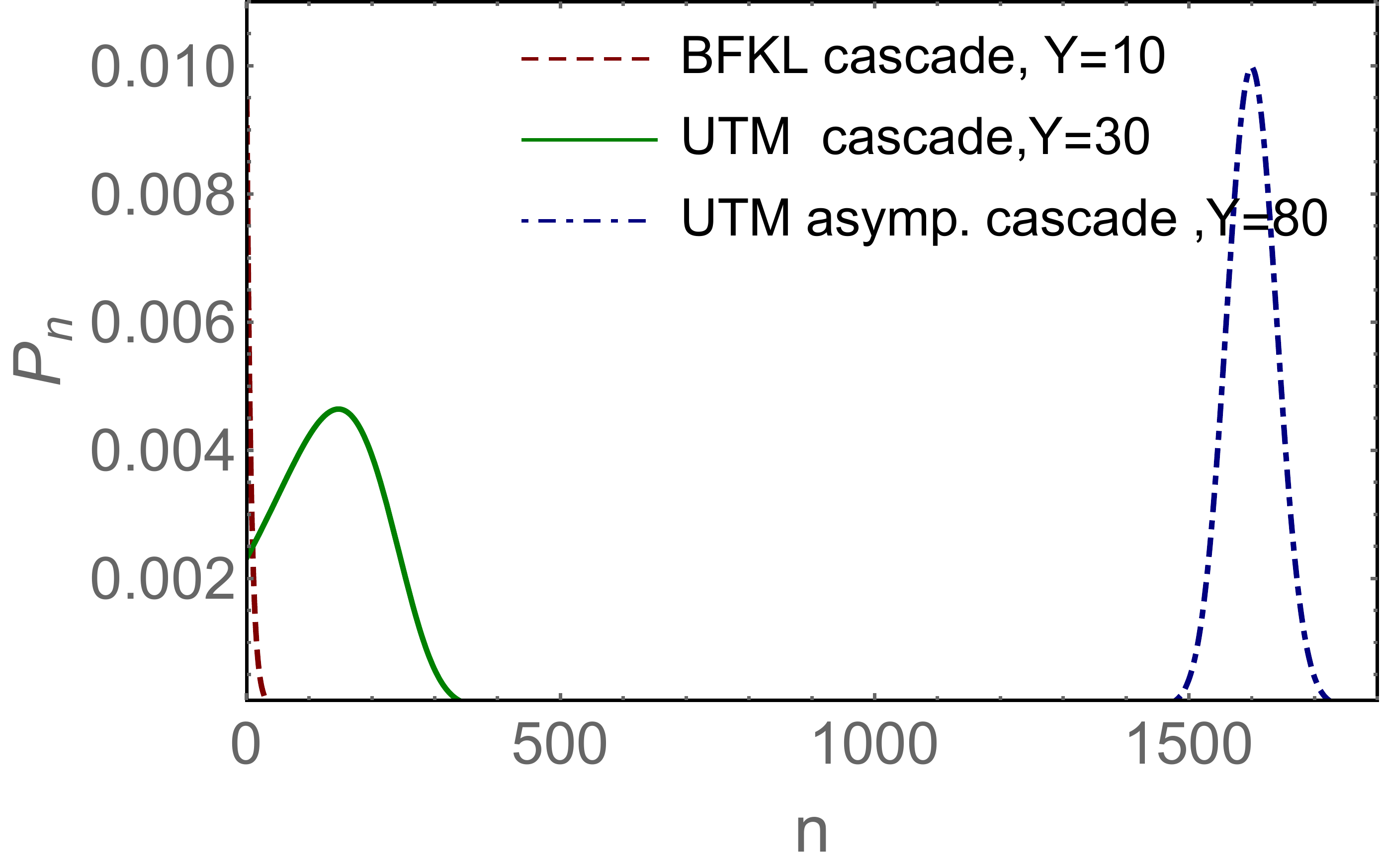}  &  \includegraphics[width=8.2cm]{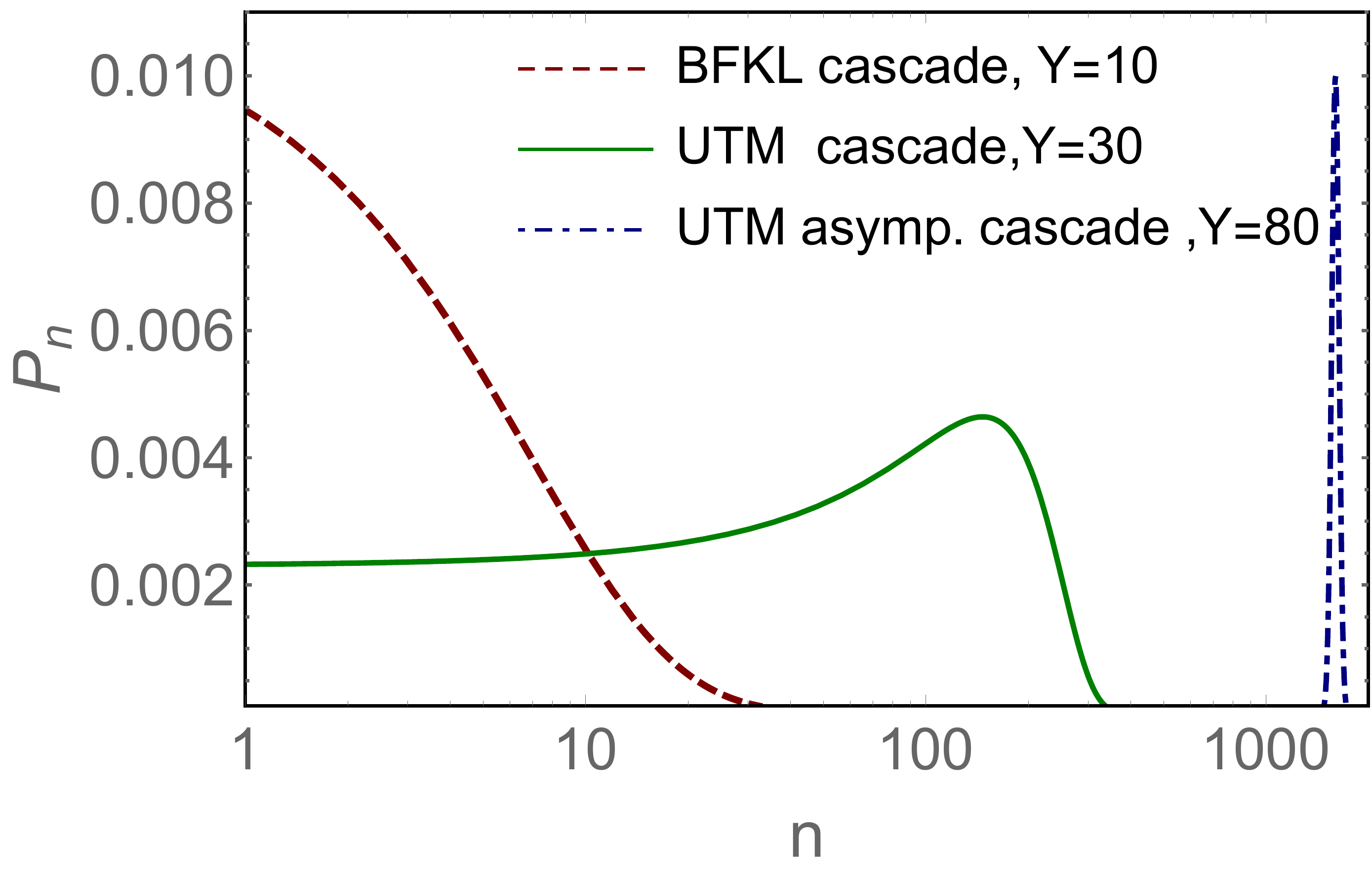}\\
      \fig{compy}-a &   \fig{compy}-b\\
            \end{tabular}
      \caption{$P_n(Y)$ versus $n$  at different $Y$ with the single dipole initial condition. $\Delta =0.2, \gamma=0.01$ }
\label{compy}
   \end{figure}

  In \fig{utmmult} we plot the dependence of the  mean multiplicity $N$ on rapidity for a single dipole initial condition, which demonstrates the above features.
  In \fig{compy} we plot the "history" of the distribution corresponding to the single dipole initial condition, starting with the BK evolution through the intermediate regime and into asymptotic rapidities.
 
  
\section{Generalizing UTM}


The UTM satisfies the physical requirements of $s$- and $t$-channel unitarity, and also provides for saturation of emission probability at very high energies. Nevertheless it is quite clear that it is incomplete as far as the description of scattering of dense objects is concerned. The main culprit here is the fact, that only a single dipole is emitted in this model in one step of the evolution. This is obvious from the Mueller-Salam derivation as well as from the evolution equation for the RFT states (\eq{evolut}). 

When the system is dense there is no reason why the number of dipoles emitted in one step of the evolution should be limited to one. Recall that the probability per unit rapidity for emission of a single dipole is of order $\Delta n$ as long as $n\le 1/\gamma$. Thus if the number of dipoles is large enough, i.e. $n\ge1/\Delta$ we expect that there is a sizable probability to emit extra dipoles, since the dipole emissions are in the first approximation independent in this range of $n$. Note that parametrically $\Delta\sim \alpha_s$ while $\gamma\sim \alpha_s^2$ thus we expect multiple dipole emissions to become important parametrically earlier than the saturation corrections.  For $n$ in the saturation regime, i.e. $n\ge 1/\gamma$ the probability density (per unit rapidity) of emission of a single dipole is actually very large $\sim 1/\alpha_s$ and multiple emissions in the wave function should be ubiquitous.

We note that in QCD at higher orders in perturbation theory the BFKL evolution indeed allows emission of more than one gluon, e.g. at NLO two gluons can be emitted. We will now discuss how multiple dipole emissions can be included in the toy model.

\subsection{Emission of two gluons  per one step of evolution}
 
 Our goal now is to include multiple dipole emissions without violating the $s$- and $t$-channel unitarity properties of the toy model. The Hamiltonian RFT formalism turns out to be a very convenient tool for this purpose. First, we already know that the $t$-channel unitarity is equivalent to the symmetry of the Hamiltonian under the transformation $H(P,\bar P)\rightarrow  H^T(\bar P,P)$. The $t$-channel unitarity is therefore very easy to implement. The $s$-channel unitarity is slightly less transparent, but with a little trial and error we can find many $s$-channel unitary evolutions. This becomes particularly simple if we consider adding to $H_{UTM}$ operators which are only (normal ordered) powers of  $\bar PP$.
 
 Consider the following simple perturbation on the Hamiltonian
  \beq \label{DELTAH}
 \delta H_\lambda \,\,=\,\,\lambda\left(\frac{\Delta}{\gamma}\right)^2 :(\bar PP)^2:\equiv \lambda\left(\frac{\Delta}{\gamma}\right)^2\bar P^2P^2
 \eeq
 where $\lambda$ is a number of order unity. The counting of the powers of $\alpha_s$ here is such that for small $n$ we recover the NLO BFKL order of magnitude of the two gluon emission. The column here denotes the normal ordered operator, meaning that all factors of $P$ have to be placed to the right of the factors of $\bar P$. The $t$-channel unitarity is clearly preserved by $\delta H_\lambda$.
 With this additional term in the Hamiltonian, the evolution equation for $P_n$ takes the form:
 
 \bea \label{DELTAH1} 
 &&\frac{\partial \,P^\lambda_n(Y)}{\partial\,Y}\,\, =\,\,  
\frac{\Delta}{\gamma}\Bigg[-\Bigg( \Lb 1 - e^{- \gamma n}\Rb \,-\, \lambda\frac{\Delta}{\gamma} \Lb 1 - e^{- \gamma n}\Rb^2\Bigg) P^\lambda_n(Y) \\
 && \,\,+\,\,\Bigg( \Lb 1 - e^{ - \gamma (n-1)}\Rb\, -\, \lambda\frac{\Delta}{\gamma} \Lb 1 - e^{ - \gamma(n - 1)}\Rb^2\Bigg) \,P^\lambda_{n-1}(Y) \,\,+\,\,\lambda\frac{\Delta}{\gamma} \Lb 1 - e^{ - \gamma(n - 2)}\Rb^2\,\,P^\lambda_{n - 2}(Y)\Bigg]\nonumber
 \eea
 For $\lambda\sim O(1)>0$ the probabilities of emission of one and two dipoles are obviously positive, and the total probability is conserved, thus this evolution is $s$-channel unitary.
 
 The RFT Schroedinger equation for the generating function now becomes
   \beq \label{DELTAH2} 
 \frac{\partial \,Z^\lambda_Y(u)}{\partial\,Y} \,\,=\,\, \frac{\Delta}{\gamma}\Bigg[ \Lb u - 1\Rb \Bigg( Z^\lambda \,\,-\,\,e^{ - \,\gamma\,u\,\frac{\partial}{\partial\,u}} \,Z^\lambda\Bigg) \,\,+\,\,\lambda\frac{\Delta}{\gamma} (1 - u)^2\,\Bigg( Z^\lambda \,\,-\,2e^{ - \,\gamma\,u\,\frac{\partial}{\partial\,u}}\,Z^\lambda +\,e^{ - 2\,\gamma\,u\,\frac{\partial}{\partial\,u}} \,Z^\lambda\Bigg)\Bigg]
 \eeq
 
 Just like for UTM, we can study the high energy asymptotics of the solution. Formally for  $\gamma u\frac{\partial}{\partial u}\gg 1$ we obtain the equation for the
asymptotic generating function as:
\beq \label{DELTAH3}
 \frac{\partial \,Z^{\lambda\, asymp}_Y ( u)}{\partial\,Y} \,\,=\,\frac{\Delta}{\gamma} \, \Bigg((u\,-\,1) \,\,+\,\,\lambda \frac{\Delta}{\gamma}\,(1\,-\,u)^2\Bigg)\,Z^{\lambda\, asymp}_Y ( u)
 \eeq
 with the solution (here, as in the previous section we take the large $Y$ limit and approximate the initial wave function at $Y_0$ by a constant):
 \beq \label{DELTAH4} 
 Z^{\lambda\, asymp} _Y(u)\,\,=\,\,\exp\Lb \,\frac{\Delta}{\gamma} \, \Bigg((u\,-\,1) \,\,+\,\,\lambda \frac{\Delta}{\gamma}\,(1\,-\,u)^2\Bigg) \,Y\Rb
 \eeq 
 From this equation we can derive $P^\lambda_n$,
 \beq\label{DELTAH5}
 P^\lambda_n\,=\,\,\oint \frac{ e^{ \,\frac{\Delta}{\gamma} \, \Bigg((u\,-\,1) \,\,+\,\,\lambda \frac{\Delta}{\gamma} \,(1\,-\,u)^2\Bigg) \,Y}}{ u^{n+1} }\,d u \,\,=\,\,\frac{1}{n!} \left[\frac{d^n}{d\, u^n} e^{ \,\frac{\Delta}{\gamma} \, \Bigg((u\,-\,1) \,\,+\,\,\lambda \frac{\Delta}{\gamma}\,(1\,-\,u)^2\Bigg) \,Y} \right]_{u=0} \eeq
 The contour of integration in \eq{DELTAH5} is the circle around  $ u =0$.  Now $P_n$ does not follow the Poisson distribution anymore.  Instead the distribution of dipoles has two independent factorial moments: $M_1=\langle n\rangle$  and   $M_2=\langle n(n-1)\rangle$: 
\begin{equation}
M^\lambda_1=\frac{\partial}{\partial u}Z^\lambda(u)|_{u=1}=\frac{\Delta}{\gamma}Y;\ \ \ M^\lambda_2 =\frac{\partial^2}{\partial u^2}Z^\lambda(u)|_{u=1}=\left(\frac{\Delta}{\gamma}\right)^2[Y^2+2\lambda Y]
\end{equation}
Clearly the additional term in the Hamiltonian modifies the asymptotic particle distribution. At first sight it may seem  that at asymptotically large rapidities these corrections  are small. Recall that the asymptotic solution is expected to be valid at such rapidities at which $
\langle n\rangle\gg 1/\gamma$, which gives $Y\gg 1/\Delta$. At these rapidities, although the magnitude of the correction to  the second moment  is large: $\delta M_2\sim \Delta/\gamma^2\sim1/\alpha^3$, the relative correction is small $\delta M_2 /M_2\sim \alpha_s$. However factorial moments are  not a very convenient measure of the shape of the distribution. If we want to get a better idea about the width of the distribution, for example we should not compare the second factorial moments but rather the variances $\sigma^2=<n^2>-<n>^2$. Here we find
\beq \sigma^2_{UTM}\xrightarrow{Y\rightarrow\infty}\frac{\Delta}{\gamma}Y;\ \ \ \ \ \ \ \ \sigma^2_\lambda\xrightarrow{Y\rightarrow\infty}\left[2\lambda\frac{\Delta^2}{\gamma^2}+\frac{\Delta}{\gamma}\right]Y;
\eeq
It is now obvious that for non-vanishing positive $\lambda$, allowing emission of the second gluon renders the asymptotic distribution much wider $\sigma^2_\lambda/\sigma^2_{UTM}\sim 1/\alpha_s$. Negative values of $\lambda$ are not physical since they lead to negative probabilities and thereby $s$-channel unitarity violation, as is obvious from \eqref{DELTAH1}.

\subsection{Emission of infinite number of gluons in  one step of evolution}
 At high density it is not realistic to think that in one step of evolution a system of partons emits only one or two  dipoles.  Rather one expects that any number of gluons can be emitted with probabilities that scale as $p_n\sim \Delta^n$ (modulo the saturation corrections). The exact values of these probabilities of course have to be determined by the underlying quantum field theory. Unfortunately in the toy world we do not have an underlying QFT,  and the appropriate toy RFT cannot be strictly derived. Nevertheless one can make a simple reasonable assumption that leads to a definite form of RFT. Let us assume that the dipoles in one step of evolution are emitted independently of each other. This is certainly what one expects if the dipole density in the wave function is large enough but the saturation corrections are still unimportant. In the language of RFT Hamiltonian a single dipole emission is represented by the operator $-\frac{\Delta}{\gamma}P\bar P$. Given that the dipoles are indistinguishable bosons, the independent emission of $n$ dipoles is represented by the operator $\frac{1}{n!}(-\frac{\Delta}{\gamma}P\bar P)^n$. Since any number of emissions $n\ge 1$ is allowed, we are lead to consider the RFT Hamiltonian 

 \beq \label{NH1}
 H_{UTMM}\,\,\,=\,\,\,:\Bigg( e^{-\frac{\Delta}{\gamma}\, P\,\bar P} \,\,-\,\,1\Bigg):
 \eeq
 where the additional M in the subscript stands for "Multiple Emissions".
 This Hamiltonian is also normal ordered so that all operators $P$ appear to the right of all the operators $\bar P$ in every order in the Taylor expansion of the exponential.
$H_{UTMM}$  reproduces UTM in the limit when $\frac{\Delta}{\gamma} \bar P P$ is small, and contains eq.(\ref{DELTAH}) with $\lambda=1/2$ as the first correction to UTM. It is obviously $t$-channel unitary.

 When acting on an $m$-dipole state it yields:
 \bea \label{HEM}
 H_{UTMM}|m\rangle&=&\left[e^{-\frac{\Delta}{\gamma}(1-e^{-\gamma m})(1-d)}-1\right]|m\rangle\nn\\
 &=&\left[e^{-\frac{\Delta}{\gamma}(1-e^{-\gamma m})}-1\right]|m\rangle+e^{-\frac{\Delta}{\gamma}(1-e^{-\gamma m})}\sum_{k=1}^\infty \frac{1}{k!}\left[\frac{\Delta}{\gamma}(1-e^{-\gamma m})\right]^k|m+k\rangle
 \eea
 This evolution of an $m$ dipole state is clearly $s$-channel unitary as well.
 
 Eq.(\ref{HEM}) is equivalent to the equations for probabilities
 \beq \label{PN1}
 \frac{dP^{\mbox{\tiny UTMM}}_n}{dY}=\left[e^{-\frac{\Delta}{\gamma}(1-e^{-\gamma n})}-1\right]P^{\mbox{\tiny UTMM}}_n+\sum_{k=1}^{\infty} \frac{1}{k!}e^{-\frac{\Delta}{\gamma}(1-e^{-\gamma (n-k)})}\left[\frac{\Delta}{\gamma}(1-e^{-\gamma (n-k)})\right]^kP^{\mbox{\tiny UTMM}}_{n-k}
 \eeq
 The Schroedinger equation for this model is
 \beq\frac{dZ^{\mbox{\tiny UTMM}}_Y(u)}{dY}=:\left[e^{-\frac{\Delta}{\gamma}(1-u)(1-e^{-\gamma u\frac{\partial}{\partial u}})}-1\right]:Z^{\mbox{\tiny UTMM}}_Y(u)
 \eeq
where the normal ordering now refers to ordering of factors $(1-u)$ and $u\frac{\partial}{\partial u}$. 

The evolution eq.(\ref{HEM}) has three distinct regimes. If the initial state contains a small number of dipoles, $m\ll 1/\Delta$, multiple emissions initially are unimportant and the cascade reduces to the BFKL-BK cascade. At higher rapidities where the typical dipole numbers are large but not "extremely large", $1/\Delta<m<1/\gamma$ one can still neglect the saturation corrections, as  $1-e^{-\gamma n}\approx \gamma n$, but multiple  emissions have to be taken into account. And finally for asymptotically large rapidities where the properties of the wave function are dominated by very large dipole number configurations $n>1/\gamma$ one has to include both multiple emissions and the saturation corrections. If one starts the evolution with a single dipole state, these regimes will successively appear as the rapidity is increased. On the other hand if initially the number of dipoles is already very large (the "large nucleus" case), the system enters the saturated regime right away. We will discuss this situation in the next section. In this section we study the probability distribution which arises in the multiple emission regime without and with saturation corrections.
\begin{boldmath}
\subsubsection{Multiple emissions without saturation ($ \frac{1}{\Delta} \,<\,n\,<\,\frac{1}{\gamma}$)}
\end{boldmath}

Let us first consider the regime where the rapidity is large enough so that multiple emissions are important, but is still too small for the saturation corrections to kick in. In terms of the dipole number $n$ this is the regime where the bulk properties of the wave function are determined by $ \frac{1}{\Delta} \,<\,n\,<\,\frac{1}{\gamma}$. The corresponding rapidity range will be given at the end of this section. In this regime we  expect UTMM to differ significantly from UTM. On the other hand at these rapidities the properties of the UTM distribution are the same as those of BFKL cascade, as discussed in the previous section. In this section we will thus compare the properties of UTMM cascade with those of BFKL, which equivalently can be thought of as a comparison between the UTMM and UTM cascades.

 For small $\gamma n$ using $1-e^{-\gamma n}\approx n\gamma$ the evolution equation for probabilities becomes
 \bea\label{smalln}
 \frac{dP^{\mbox{\tiny UTMM}}_n}{dY}&=&\left[ e^{-\Delta n}-1\right]P_n^{\mbox{\tiny UTMM}}+\sum_{k=1}^{\infty} \frac{1}{k!}e^{-\Delta(n-k)}\left[\Delta(n-k)\right]^kP^{\mbox{\tiny UTMM}}_{n-k}\,\nn\\\,&=&\,\,-P^{\mbox{\tiny UTMM}}_n\,+\,\sum_{k=0}^{\infty} \frac{1}{k!}e^{-\Delta(n-k)}\left[\Delta(n-k)\right]^kP^{\mbox{\tiny UTMM}}_{n-k}\eea
We note that this equation is consistent with normalization of probability, as summing \eq{smalln} over $n$ we find that $\frac{d}{dY}\sum_{n} P_n^{\mbox{\tiny UTMM}}=0$.

The Schroedinger equation in this limit becomes
 \beq \label{SE1}
 \frac{dZ^{\mbox{\tiny UTMM}}_Y(u)}{dY}=:\left[e^{-\Delta(1-u) u\frac{\partial}{\partial u}}-1\right]:Z^{\mbox{\tiny UTMM}}_Y(u)
 \eeq
 Just like in the BK limit, the no-saturation limit of UTMM is not sensitive to the value of $\gamma$ - the only constant that enters here is the emission probability $\Delta$.
 To study the properties of the resulting probability distribution, we rewrite \eq{smalln} as evolution equation for the factorial moments.
Let us start with the mean multiplicity $M_1^{\mbox{\tiny UTMM}}$. \eq{smalln} takes the form
\begin{eqnarray} \label{M1}
 \frac{d M_1^{\mbox{\tiny UTMM}}}{dY}&=& - M_1^{\mbox{\tiny UTMM}}\,+\,\sum_{k=0, n=1}^{\infty}\Big( n - k + k\Big) \frac{1}{k!}e^{-\Delta(n-k)}\left[\Delta(n-k)\right]^kP^{\mbox{\tiny UTMM}}_{n-k}\nonumber \\
 &=&\,- M_1^{\mbox{\tiny UTMM}}\,+ \Delta M_1^{\mbox{\tiny UTMM}} \,\,+\,\,M_1^{\mbox{\tiny UTMM}} =\,\Delta\,M_1^{\mbox{\tiny UTMM}}
 \end{eqnarray}
 We consider the initial conditions corresponding to a single dipole initial state $M_k^{\mbox{\tiny UTMM}}(Y=0)=\delta_{k1}$.
 With this initial condition the solution is
 \beq \label{M11}
 M_1^{\mbox{\tiny UTMM}}\Lb Y\Rb\,\,=\,\,e^{\Delta\,Y}
 \eeq
 Interestingly the mean multiplicity turns out to be the same as for BK distribution (see \eq{PNBK}).
 For $M_2^{\mbox{\tiny UTMM}}$ we have
 \bea \label{M2}
 \frac{d M_2^{\mbox{\tiny UTMM}}}{dY} &=& - M_2^{\mbox{\tiny UTMM}}\,+\,\sum_{k=0, l=1}^{\infty}\Big( (k + l)^2 - (k+l) \Big) \frac{1}{k!}e^{-\Delta(l)}\left[\Delta(l)\right]^kP^{\mbox{\tiny UTMM}}_{l}\\
  &=&- M_2^{\mbox{\tiny UTMM}}\,+ \,M_2^{\mbox{\tiny UTMM}}\,+\, \Lb 2\,\Delta\,+\,\Delta^2\Rb\Lb M_2^{\mbox{\tiny UTMM}}\,+M_1^{\mbox{\tiny UTMM}}\Rb\,=\,\Lb 2\, \Delta\,+\Delta^2\Rb \Lb M_2^{\mbox{\tiny UTMM}}\,+\,M_1^{\mbox{\tiny UTMM}}\Rb \nn\eea
  For a single dipole initial condition the solution for $M_2^{\mbox{\tiny UTMM}}$:
  \beq \label{M21}
 M_2^{\mbox{\tiny UTMM}}\Lb Y\Rb\,\,=\,\,\frac{ (2 + \Delta)}{  (\Delta + 1)}\Bigg( e^{( 2  + \Delta)\Delta\,Y} \,\,-\,\,e^{\Delta\,Y}\Bigg)
 \eeq  
 For $M_3^{\mbox{\tiny UTMM}}$ the equation is:
 \bea \label{M3}
 \frac{d M_3^{\mbox{\tiny UTMM}}}{dY} &=& - M_2^{\mbox{\tiny UTMM}}\,+\,\sum_{k=0, l=1}^{\infty}\Big( (k + l)^3 - 3(k+l)^2 + 2(k +l) \Big) \frac{1}{k!}e^{-\Delta(l)}\left[\Delta(l)\right]^kP^{\mbox{\tiny UTMM}}_{l}\,\nn\\
  &=&\,\,- M_3^{\mbox{\tiny UTMM}}\,+ \,M_3^{\mbox{\tiny UTMM}}\,\,+\,\, \Lb  ( 1\,+\, \Delta)^3 - 1)\Rb\Lb M_3^{\mbox{\tiny UTMM}} \,+\,3\,M_2^{\mbox{\tiny UTMM}} \,+M_1^{\mbox{\tiny UTMM}}\Rb\nn \\ 
  &-&\,3 \,\Delta\Lb M_2^{\mbox{\tiny UTMM}}\,+\,M_1^{\mbox{\tiny UTMM}}\Rb\,\nn\\
  &=& \Lb  ( 1\,+\, \Delta)^3 - 1)\Rb  \,M_3^{\mbox{\tiny UTMM}} \,\,+\,\,3\,\Delta (1 + \Delta)\,(2 \,+\,\Delta)\,M_2^{\mbox{\tiny UTMM}}\,\,+\,\,\Delta^2\,(3 + \Delta)\,M_1^{\mbox{\tiny UTMM}}
  \eea 
  with the solution
  \beq \label{M31}
  M_3^{\mbox{\tiny UTMM}} =\frac{e^{\Delta  Y} \left((\Delta +1) (\Delta  (2 \Delta +9)+12)-3 (\Delta +2)^3 e^{\Delta  (\Delta +1) Y}+(\Delta +4) (\Delta  (\Delta +3)+3) e^{\Delta  (\Delta +1) (\Delta +2) Y}\right)}{(\Delta +1)^2 (\Delta +2)} 
  \eeq
   Considering $\Delta\,\ll\,1$ we can simplify \eq{M21} and \eq{M31} (we do not neglect corrections due to nonvanishing $\Delta$ in the exponent, but only in the prefactors):
  \beq \label{M32}
 M_2^{\mbox{\tiny UTMM}}\Lb Y\Rb\,\,=\,\,2!\Bigg( e^{((1+ \Delta)^2-1)\,Y}  \,\,-\,\,e^{\Delta\,Y}\Bigg) ; \ \ \ \ \ \ \ 
   M_3^{\mbox{\tiny UTMM}}\Lb Y\Rb\,\,=\,\,3!\Bigg( e^{((1+ \Delta)^3-1)\,Y} \,\,-\,\,2 \,e^{((1+ \Delta)^2-1)\,Y}\,\,+\,\,e^{ \Delta\,Y}\Bigg)
   \eeq   
    For $\Delta^2 Y \,\ll\,1$
 we see that $M_2^{\mbox{\tiny UTMM}} \,\approx\, 2 M_1^{\mbox{\tiny UTMM}}
 \Big( (M_1^{\mbox{\tiny UTMM}})^{1+\Delta}\,-\,1\Big)$ and the second moment of UTMM is very close to that of the BK multiplicity distributions of \eq{PNBK}. The same holds for the third moment (neglecting $\Delta$ in the exponent) $M_3^{\mbox{\tiny UTMM}} \approx 3!\,M_1^{\mbox{\tiny UTMM}}\,\Lb M_1^{\mbox{\tiny UTMM}}\,-\,1\Rb^2 $. At much larger rapidities  $\Delta^2 Y \,>\,1$ the difference between the moments in  UTMM and BK is significant, and the above expressions for both
 $M_2^{\mbox{\tiny UTMM}}$ and $M_3^{\mbox{\tiny UTMM}}$ are much larger than in the BK limit. 
 However we should keep in mind  that the above expressions for $M_{k}^{\mbox{\tiny UTMM}}$ are representative of the UTMM only for $M_1^{\mbox{\tiny UTMM}}<1/\gamma$, which translates into $Y<\frac{1}{\Delta}\ln \frac{1}{\gamma}$. For rapidities $Y\gg 1/\Delta^2\sim 1/\gamma$ the saturation corrections take over and the actual distribution is not described faithfully by the solution to eq.(\ref{smalln}). Thus rapidities $Y\sim \frac{1}{\Delta^2}$ are strictly speaking outside the range of validity of our present approximation. 
 
\begin{figure}[ht]
    \centering
  \leavevmode
      \includegraphics[width=10cm]{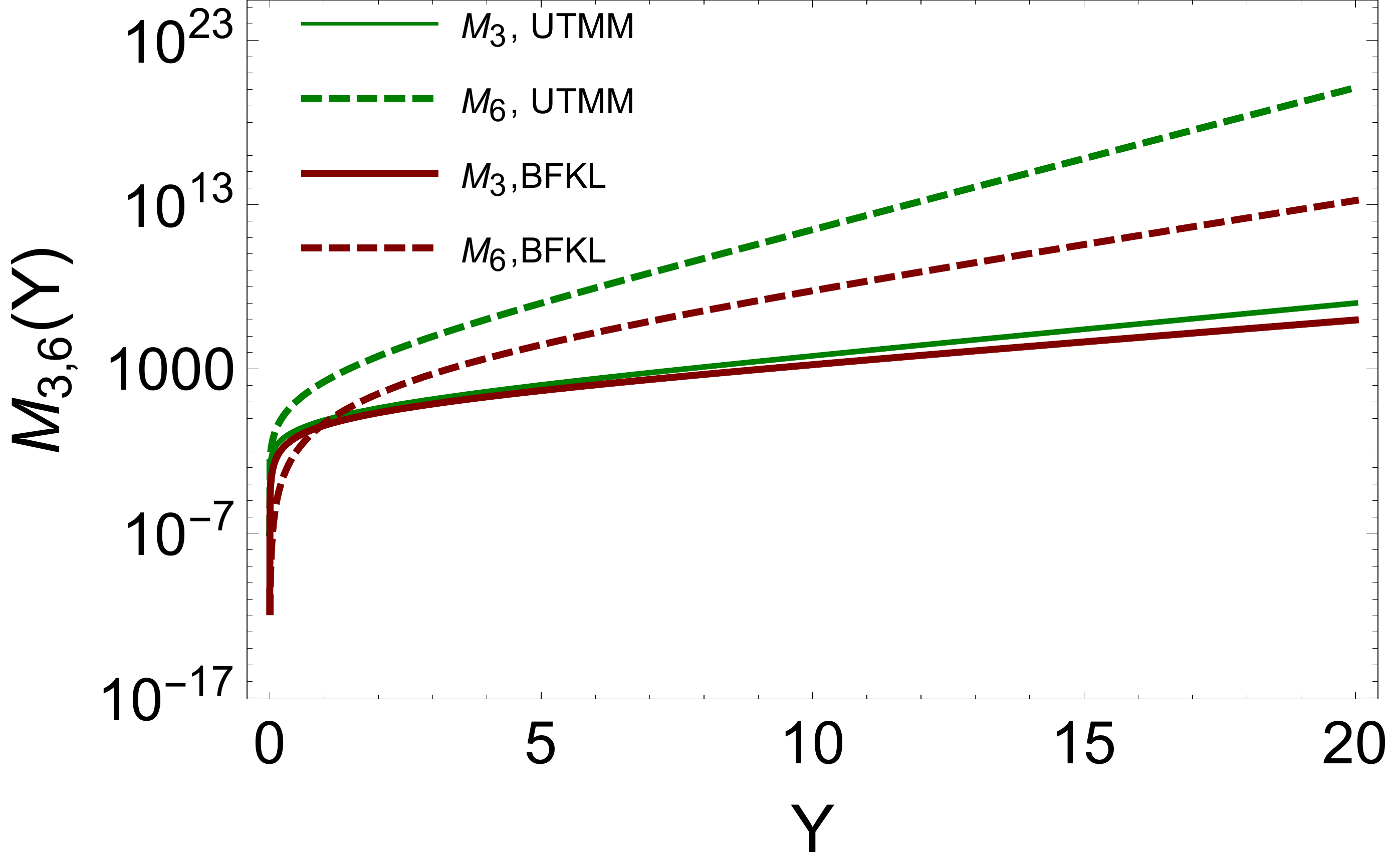}  
      \caption{ $M_k$   versus $Y$. The green curves describe the exact solution to \eq{M31} for the UTMM cascade and \eq{MGF4}, while the red ones  correspond to the BFKL   multiplicity distributions of \eq{MKBK}. $\Delta = 0.2$.}
\label{mmd}
   \end{figure}

 Thus for rapidities that we may consider, the expressions in eqs.\eqref{M21},\eqref{M32} are (up to perturbative corrections) the same as in the BFKL -BK regime. This at the first glance looks strange, since the UTM allows emissions of multiple dipoles in one step of the evolution. The reason we do not get on  average more dipoles in the wave function, as well as nearly identical second and third moments of the distribution, is because the exponential form of the Hamiltonian includes a kind of "unitarization corrections", i.e. since many dipoles are allowed to be emitted, the probability of emission of a single dipole is smaller than in the BK model (at the same value of parameter $\Delta$). Nevertheless even in the allowed region we expect the distribution to differ significantly from BK. Indeed, examining higher moments we see that this is the case. Let us consider only the leading term in the $k$-th moment, i.e. the term with the fastest growth in rapidity. For this term it is easy to write the general expression ( see Appendix A)
      \beq
   M_k^{\mbox{\tiny UTMM}}(Y)\propto e^{((1+\Delta)^k-1)Y}
   \eeq
   Consider high moments, so that $k=\frac{K}{\Delta}$. For very small $\Delta$ we have $(1+\Delta)^{\frac{1}{\Delta}}\approx e$, and thus
   \beq
   M_k^{UTMM}(Y)\propto e^{(e^K-1)Y}
   \eeq
   while for the same $k$ the leading behavior in the BK model is
   \beq
   M_k^{BK}\propto e^{KY}
   \eeq
   Thus for large $k$ the moments differ already in the leading order term with
   \beq
   M_k^{UTMM}\gg M_k^{BK}
   \eeq
   We conclude that, just like expected the probability distribution is significantly wider in the "multiple emission" regime of UTMM compared to BFKL-BK cascade.
   In \fig{mmd} we plot the third and sixth moments for illustration and indeed see that while $M_3$ in both cascades is very similar at intermediate rapidities, $M_6$ is significantly larger in UTMM.

 In principle, once the factorial moments are given, one can reconstruct the probability distribution. In the present case this turns out to be tricky since the moments grow very fast at large $k$ and one is faced with a necessity to sum an asymptotic series. In Appendix A we discuss a way to resolve this issue and give a derivation of the (approximate) probability distribution. The result is 
 \beq \label{MGFPN1}
P^{\mbox{\tiny UTMM}}_n\Lb Y\Rb  =\,\,e^{-Y}\sum^{\infty}_{j=0} \frac{Y^j}{j!} \,P^j_n~~~~~\mbox{with}~~~
P^j_n\,\,=\,\,\frac{1}{N_j} \Lb 1 + \frac{1}{N_j}\Rb^{-n }\,
 \eeq  
 where $N_j\,\,=\,\,\Lb1\,+\,\Delta\Rb^j$.

\begin{figure}[ht]
    \centering
  \leavevmode
  \begin{tabular}{cc}
      \includegraphics[width=8.6cm]{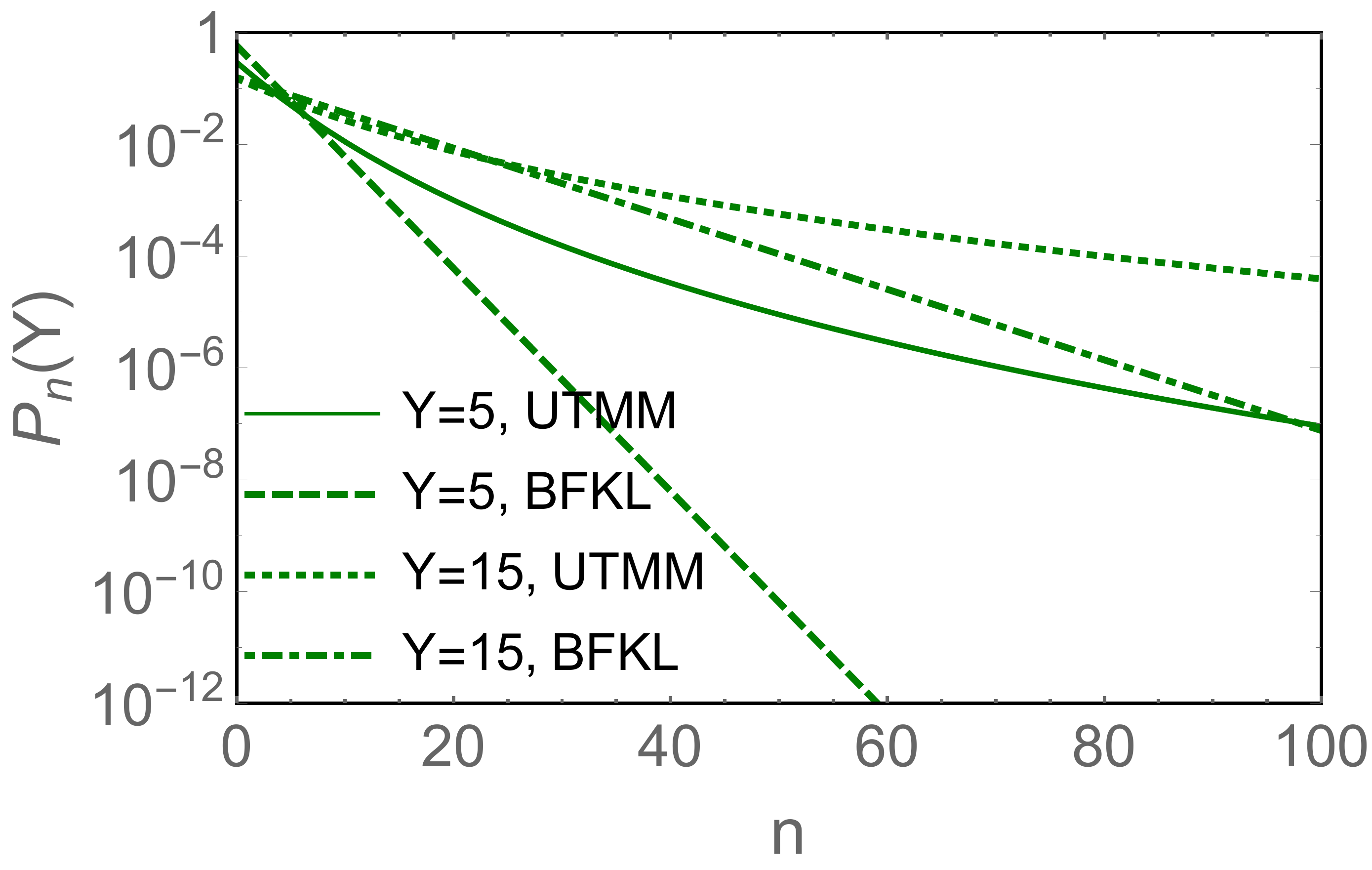} &  \includegraphics[width=8.4cm]{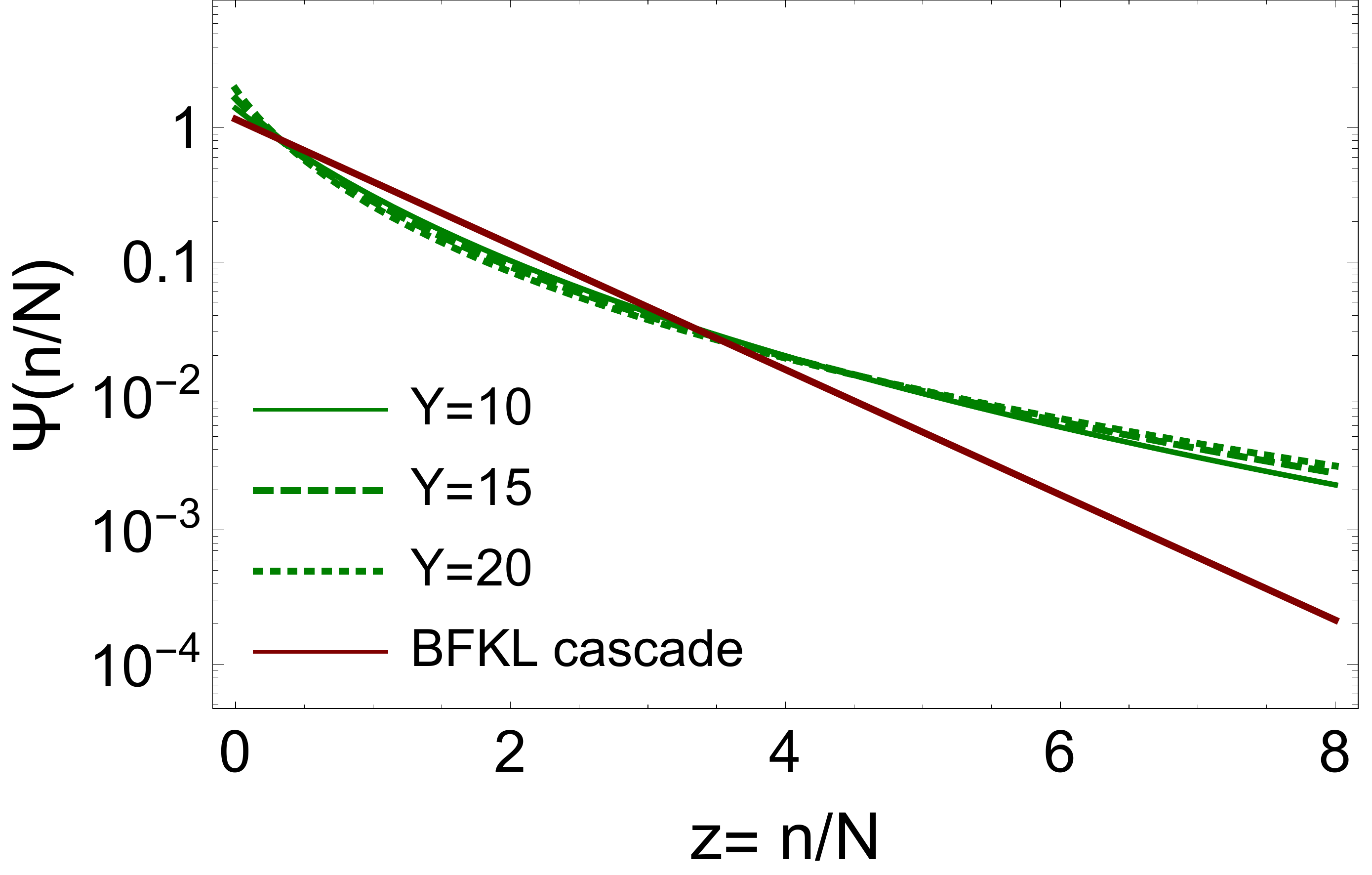}  \\
      \fig{pnn}-a & \fig{pnn}-b\\
      \end{tabular}
          \caption{\fig{pnn}-a: $P_n(Y)$ versus $n$ at fixed values of Y  from \eq{MGFPN1}. \fig{pnn}-b: the KNO function $\Psi\Lb z\Rb\,\,=N P_{z\,N} $ where $N$ is the mean multiplicity. $\Delta = 0.2$. BFKL cascade  denotes the distribution of \eq{PNBK}.
}
\label{pnn}
   \end{figure}

 In \fig{pnn} we show the values of $P^{UTMM}_n $ from \eq{MGFPN1} versus  $n$ at different values of $Y$. 
 \fig{pnn}-a shows that  in the region of small $n$ the UTMM multiplicity distribution is close to BFKL cascade   but at large $n$ the deviation  is rather large. This is consistent with our discussion of high factorial moments.  \fig{pnn}-b shows that although UTMM cascade strictly speaking violates KNO scaling, these violations  are not large in practice.

 Finally we note that we can estimate the range of rapidities in which the evolution is dominated by multiple emissions, but saturation is still not important. The relevant condition in terms of the average mulitplicity is $1/\Delta< M_1^{\mbox{\tiny UTMM}}<1/\gamma$, which given \eq{M11} in terms of rapidity translates into $\frac{1}{\Delta}\ln\frac{1}{\Delta}<Y<\frac{1}{\Delta}\ln \frac{1}{\gamma}$.
 \subsubsection{The S-matrix for single and multiple gluon emissions}
 
 Although our main focus in this paper is in the probability distributions, it is interesting to see how allowing multiple emission affects more directly measurable physical quantities.
 
The S-matrix  can be calculated  using \eqref{S} \cite{MUSA}. For UTM in the kinematic regime under consideration we can use the probabilities given in \eq{PNBK1}
 \bea \label{SMS1}
 S(Y) \,\,&=&\,\,\sum^\infty_{n,m=0} e^{ - m \,n\,\gamma} \,P^{\mbox{\tiny BFKL}}_n( Y_0)\,P^{\mbox{\tiny BFKL}}_m( Y - Y_0)\\
 & =&\,\,\int_0^\infty \frac{d n}{n} \,d\,u\, e^{-\gamma\,u}\frac{1}{  N(Y_0) \,N(Y - Y_0)} \,\exp\Lb - \frac{n}{N(Y_0)} \,-\,
\frac{u}{ n\,N(Y \,-\,Y_0)} \Rb\,\,
 =\,-\frac{e^{\frac{1}{\gamma  N(Y)}} \text{Ei}\left(-\frac{1}{N(Y) \gamma }\right)}{\gamma  N(Y)} 
 \nn 
\eea
where $Ei(z)$ is the exponential integral: $ Ei(z) = - E_1(z)=\int\limits^z_{-\infty}\frac{e^\xi}{\xi} d \xi$.

For the total cross section we have\footnote{ This is a slight abuse of language, as in the toy model there is no transverse dimension, and thus no cross section. The quantities we calculate in this subsection are dimensionless, and are simply proportional to the appropriate probabilities. We will nevertheless refer to them as cross sections using the higher dimensional analogy.}
 \beq \label{COM10}
  \sigma_{tot}\,\,=\,\,2 \,A(Y)= 2\Lb 1\,-\,S(Y)\Rb
  \eeq
  We can also  estimate the inelastic cross section, using the approach of \cite{MUSA}
  \beq \label{COM2}
  \sigma_{in}\,=\,1\,\,-\,\,\,\,\sum_{n,m} e^{ -2\, m \,n\,\gamma} \,P_n( Y_0)\,P_m( Y - Y_0)    \eeq
  In general the inelastic cross section defined in \eq{COM2} is not independent of frame, i.e. depends on $Y_0$ \cite{MUSA}. However for the probability distribution  \eq{PNBK1} we find an $Y_0$-independent result
 \beq \label{COM3} 
  \sigma_{in}\,\,=\,\,A\Lb Y,\gamma \to 2 \gamma\Rb\,\,=\,\,1\,\,+\frac{e^{\frac{1}{2\,\gamma  N(Y)}} \text{Ei}\left(-\frac{1}{2\,\gamma\,N(Y) }\right)}{2\,\gamma  N(Y)}\eeq
With $\sigma_{tot} $ and $\sigma_{in}$ we can estimate the cross section for diffraction production ( $\sigma_{diff}$):
  \beq \label{COM4}
  \sigma_{diff}\,\,=\,\,\sigma_{tot}\,\,-\,\,\sigma_{in}
  \eeq
  With this definition  $\sigma_{diff}$ includes the elastic cross section.
  The cross section for diffractive dissociation is obtained by subtracting for $\sigma_{diff}$ the elastic cross section
  \beq\label{COM5}
  \sigma_{dd}=\sigma_{diff}-(A(Y))^2
  \eeq

For UTMM, replacing $P_n$ in \eq{SMS1} by \eq{MGFPN1} we obtain 
 \bea \label{SMS2}
 S^{\mbox{\tiny UTMM}} (Y) \,\,&=&\,\,\sum_{n,m} e^{ - m \,n\,\gamma} \,P^{\mbox{\tiny UTMM}}_n( Y_0)\,P^{\mbox{\tiny UTMM}}_m( Y - Y_0)\\
 & =&\,\,e^{-Y}\sum_{j_1} \sum_{j_2}\frac{Y_0^{j_1} \,\Lb Y - Y_0\Rb^{j_2}}{j_1!\,j_2!}  \int_0^\infty \frac{d n}{n} \,d\,u\, e^{-\gamma\,u}\frac{1}{  N_{j_1}N_{j_2}} \,\exp\Lb - \frac{n}{N_{j_1}} \,-\,
\frac{u}{ n\,N_{j_2}} \Rb\,\,
 \nn \\
&=&\,\,-\,e^{-Y} \sum_{j} \frac{Y^{j}}{(j)!} \frac{e^{\frac{1}{\gamma  N_{j}}} \text{Ei}\left(-\frac{1}{\gamma\,N_{j} }\right)}{\gamma \,N_{j}}\nn
 \eea

    Using  \eq{SMS2} in \eq{COM10}-\eq{COM5} we obtain the physical observables in the UTMM.

        \begin{figure}[h]
    \centering
  \leavevmode
    \includegraphics[width=9cm]{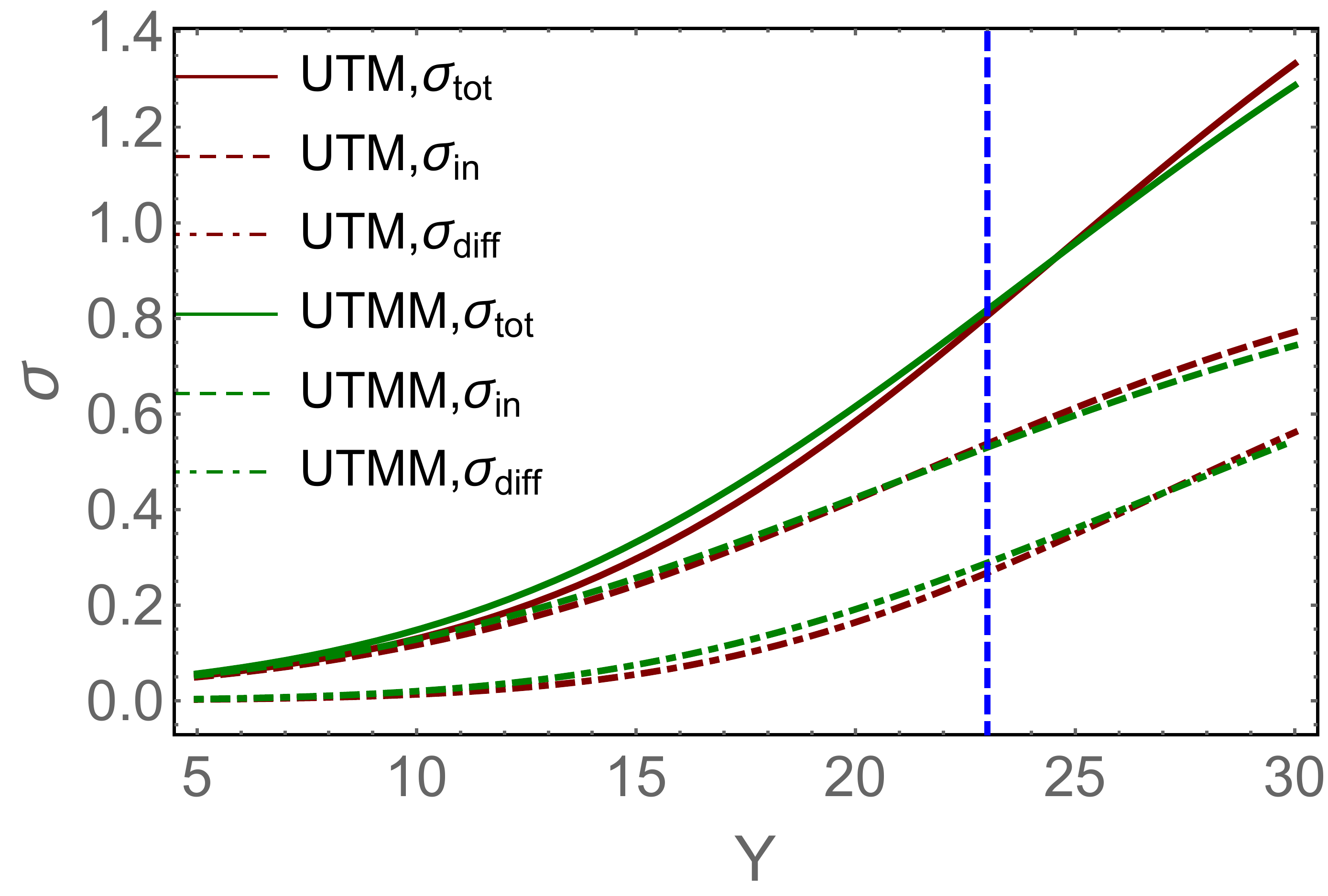}     
   \caption{ The inclusive observable  in UTM and UTMM  versus Y. $\Delta=0.2$  and $\gamma = 0.01$. 
       }
\label{comxs1}
   \end{figure}
       \begin{figure}[h]
    \centering
 \leavevmode
    \includegraphics[width=9cm]{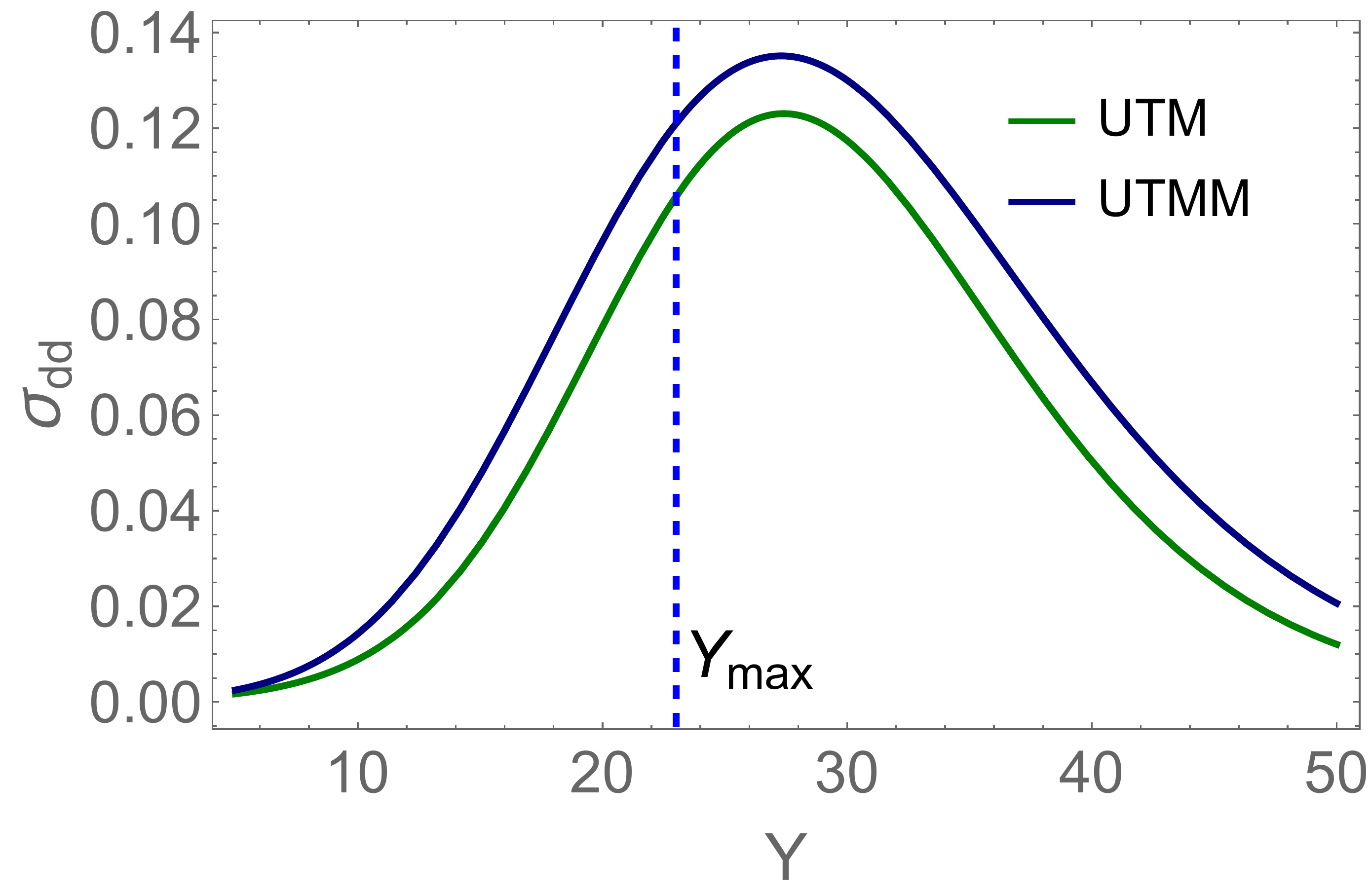}     
   \caption{ The cross section of diffraction dissociation  $\sigma_{dd} = \sigma_{diff} - \sigma_{el}$ from  \eq{COM4}, for UTM and UTMM   versus Y. $\Delta=0.2$  and $\gamma = 0.01$. 
       }
\label{div1}
  \end{figure}

      In \fig{comxs1} and \fig{div1} we plot the various cross sections for the two models (UTM and UTMM). The kinematic region where saturation corrections are not important extends only up to rapidity $Y_{\rm max}= \frac{1}{\Delta}\ln \frac{1}{\gamma}$. denoted by the vertical line on the graphs. We nevertheless plot the curves up to higher rapidities to illustrate an interesting point that indeed around $Y_{max}$ the various quantities exhibit qualitative change of behavior. The differences between the two models follow the expected trend. The total cross section in UTMM is higher than in UTM, since it has larger  probability to have higher number of dipoles. The cross section for diffractive dissociation on the other hand is lower in UTMM, consistent with the fact that it involves subtraction of the square of elastic amplitude, which is sensitive to the fluctuation in the dipole number.
   In fact  one can see from the graphs that among the quantities we calculated, $\sigma_{dd}$  is the best discriminator betweem UTM and UTMM.

\begin{boldmath}
\subsubsection{Multiple emissions with saturation - large $Y$ asymptotics ($ \,n\,\gg\,\frac{1}{\gamma}$)}
\end{boldmath}


For very large $n$  the term $\exp(-\,\gamma n)\,\,\ll\,1$  can be neglected and equations for $P_n$ take the form:
 \beq \label{NMN1}
 \frac{d \,P^{\mbox{\tiny UTMM}}_n(Y)}{  d\,Y} \,\,=\,\, \Lb  e^{ - \frac{\Delta}{\gamma}} \,-\,1\Rb P^{\mbox{\tiny UTMM}}_n(Y) \,+\, e^{ - \frac{\Delta}{\gamma}}  \,\sum_{k=1}^{\infty} \frac{1}{k!}\Lb \frac{\Delta}{\gamma}\Rb^k\,P^{\mbox{\tiny UTMM}}_{n - k}(Y)
 \eeq
  and the asymptotic form of the RFT Schroedinger equation is:
 
 \beq \label{NH4}
  \frac{d\,Z^{\mbox{\tiny UTMM}}_{asymp}\Lb Y\Rb}{ d\, y}\,\,=\,\, \Bigg( e^{ \frac{\Delta}{\gamma}\Lb u  -1\Rb} \,\,-\,\,1\Bigg)\,Z^{\mbox{\tiny UTMM}}_{asymp}\Lb Y\Rb  \eeq
  This has an  obvious solution:
  \beq \label{NH5}
  Z^{\mbox{\tiny UTMM}}_{asymp}\,\,=\,\,\exp \Lb\Bigg( e^{ \frac{\Delta}{\gamma} \Lb u  -1\Rb} \,\,-\,\,1\Bigg)\,\,Y\Rb
  \eeq  
 where, as before we have neglected a possible slow varying prefactor $Z_{Y_0}(u)$ arising from initial condition.
  These equations describe the asymptotics of the distribution in UTMM at very large $Y$.
 
 The factorial moments are obtained using \eq{DELTAH5} and the asymptotic solution of \eq{NH5}.  
 For the first three factorial moments we then obtain
 \beq \label{NH7}
 M_1^{\mbox{\tiny UTMM}} = \frac{\Delta}{\gamma} Y;~~~~~ M_2^{\mbox{\tiny UTMM}}= \frac{\Delta^2}{\gamma^2} Y^2  + \frac{\Delta^2}{\gamma^2} Y;~~~~
  M_3^{\mbox{\tiny UTMM}}= \frac{\Delta^3}{\gamma^3} Y^3\,+3\,\frac{\Delta^3}{\gamma^3} Y^2 \,+\, \frac{\Delta^3}{\gamma^3} \,Y\, 
 \eeq
 One can write the moments in the following explicit form 
 \beq \label{NH8}
M_k^{\mbox{\tiny UTMM}}\,\,=\,\, \Lb \frac{\Delta}{\gamma}
t\frac{d}{d\,t}\Rb^k e^{(t - 1)\,Y}|_{t=1}
 \eeq
 with
 \beq
 t\equiv  \exp\Lb \frac{\Delta}{\gamma}(u-1) \Rb
 \eeq
 Similar representation for the probabilities is 
 \beq \label{NH9}
  P^{\mbox{\tiny UTMM}}_n\Lb Y\Rb\,\,=\frac{1}{n!}\, \Lb \frac{\Delta}{\gamma}
t\frac{d}{d\,t}\Rb^n e^{(t - 1)\,Y}|_{t=e^{-\frac{\Delta}{\gamma}}}\eeq
Another representation can be obtained   expanding the exponent $e^{t\,Y }$:
  \bea \label{NH10}
  P^{\mbox{\tiny UTMM}}_n\Lb Y\Rb\,\,&=&\frac{1}{n!}\, \Lb \frac{\Delta}{\gamma}
t\frac{d}{d\,t}\Rb^n e^{(t - 1)\,Y}|_{t=e^{-\frac{\Delta}{\gamma}}}
=\,\,\frac{1}{n!}\, e^{-Y}\Lb \frac{\Delta}{\gamma}
t\frac{d}{d\,t}\Rb^n \sum_{k=0}^\infty \frac{\Lb Y\,t\Rb^k}{k!} |_{t=e^{-\frac{\Delta}{\gamma}}} \nn\\
&=&\, \,\frac{1}{n!}\, e^{-Y} \sum_{k=0}^\infty \frac{1}{k!}\left(\frac{\Delta}{\gamma}k\right)^n\left(Ye^{-\frac{\Delta}{\gamma}}\right)^k
  \eea

  An interesting question is what is the range of $k$ that contributes to the sum in \eq{NH10}. For large rapidity we expect that the most important values of $n$ are large, and $k$ for these $n$ is also large. For large $k$ we can estimate the range
by replacing the sum with the integral over $k$ and finding the maximum of the integrand. The equation for the saddle point, $k_{SP}$ has the form:
  \beq \label{NH11}
  \frac{d \Psi}{d k}|_{k=k_{SP}} =  0~~~\mbox{with}~~\Psi =  n\Lb \ln\Lb \frac{\Delta}{\gamma}\Rb + \ln \Lb k\Rb\Rb\,-\,k\,\Lb \ln\Lb \frac{k}{\cal Y}\Rb - 1\Rb;
 ~~~~~~ \frac{n}{k_{SP}} \,-\,\ln\frac{ k_{SP}}{\cal Y}=0;
  \eeq
  where ${\cal Y} = Y\exp\Lb - \frac{\Delta}{\gamma}\Rb$.  The approximate solution  for $k_{SP}$ is
  \beq \label{KSP}
  k_{SP}\,\,=\,\,\frac{n}{\ln\Lb \frac{n}{\cal Y}\Rb}
  \eeq
  Note that $e^{-\frac{\Delta}{\gamma}}$ is an exponentially small number, thus for reasonable values of $Y$ we have ${\cal Y}\ll 1$. Hence $k_{SP}\ll n$ and the ratio between the two decreases for large $n$.
  
\begin{figure}[h]
      \centering
  \leavevmode  \begin{tabular}{c c}
    \includegraphics[width=8cm]{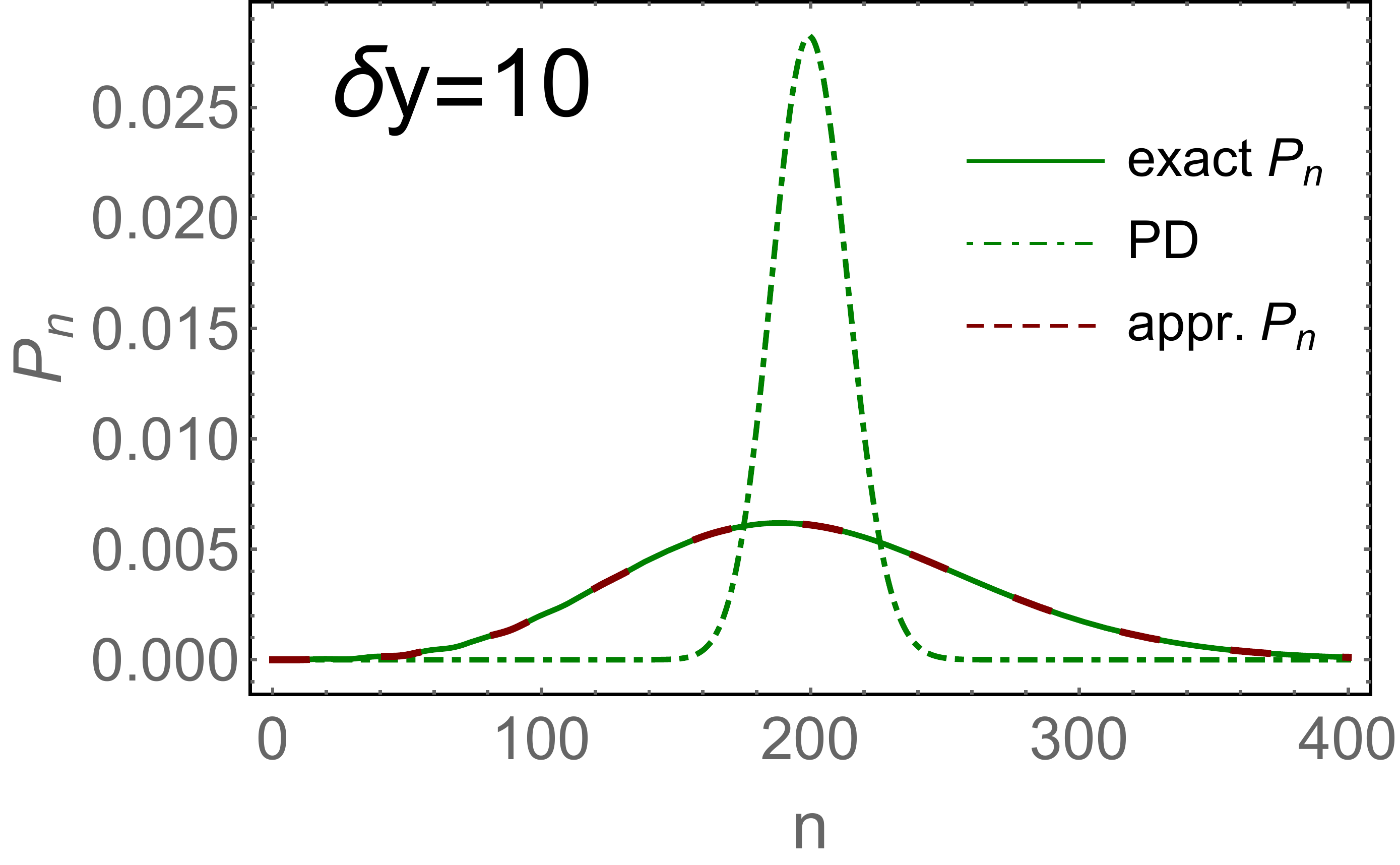}  & \includegraphics[width=8cm]{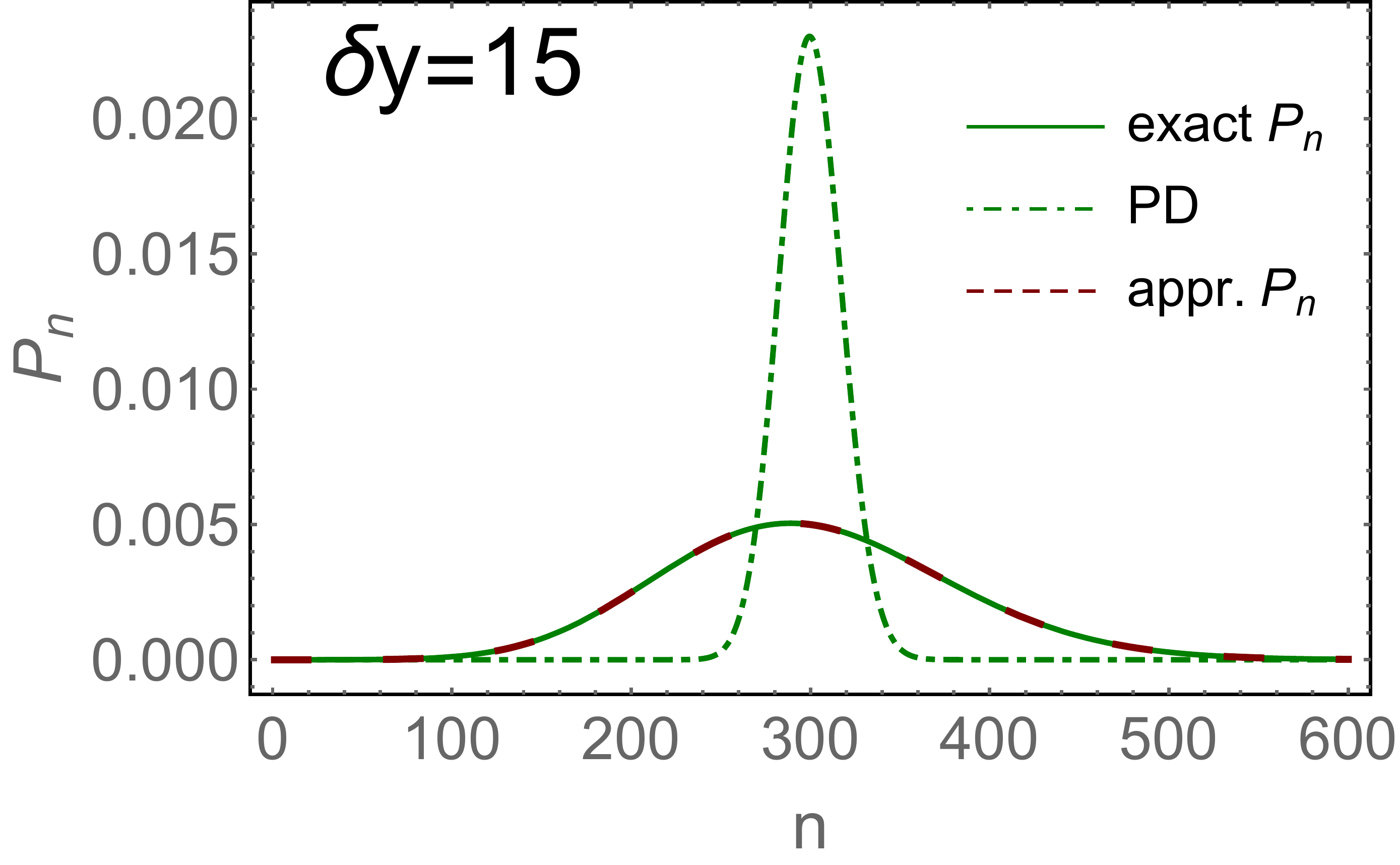}  \\
    \fig{md}-a & \fig{md}-b\\
     \includegraphics[width=8cm]{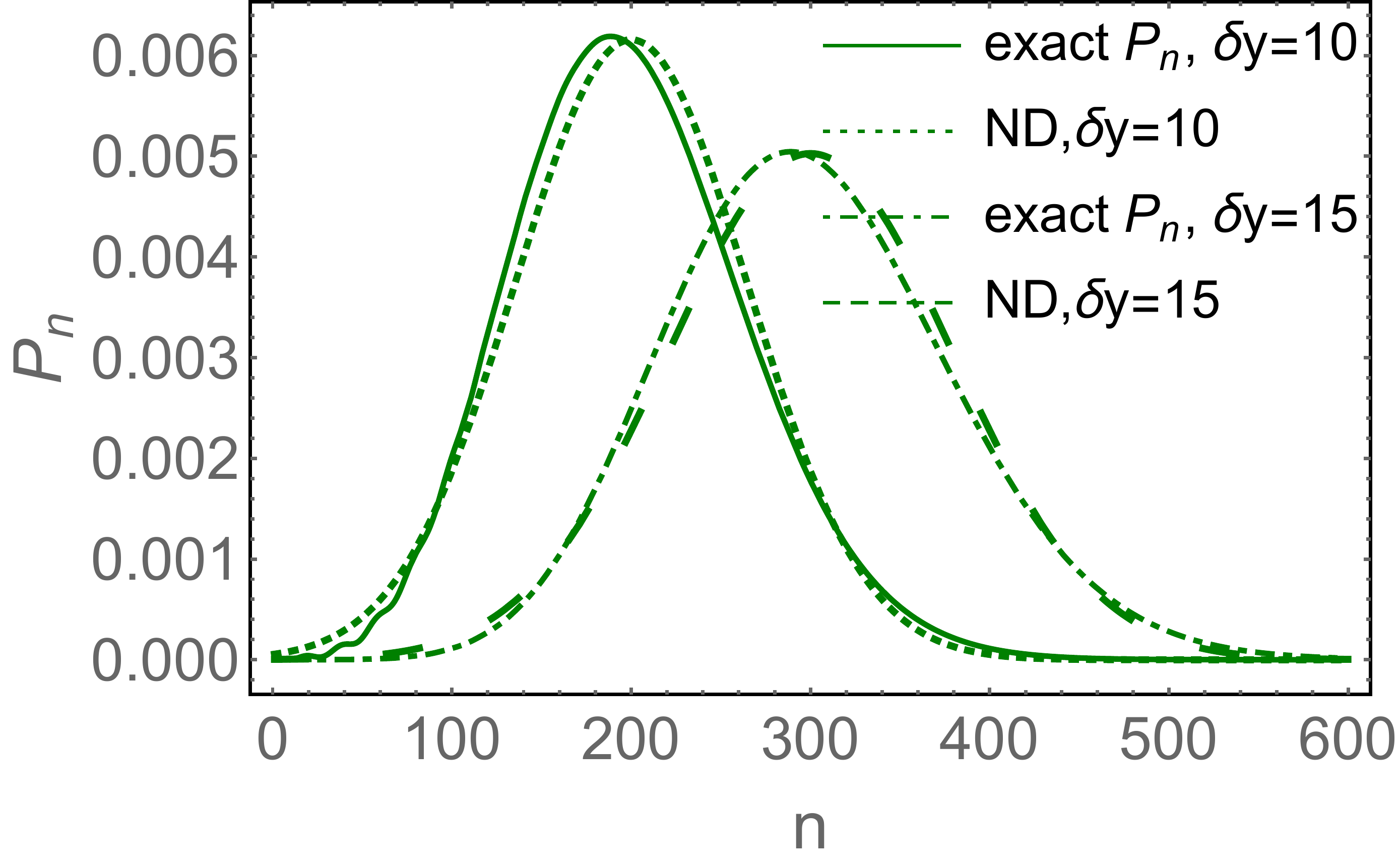}  & \includegraphics[width=8cm]{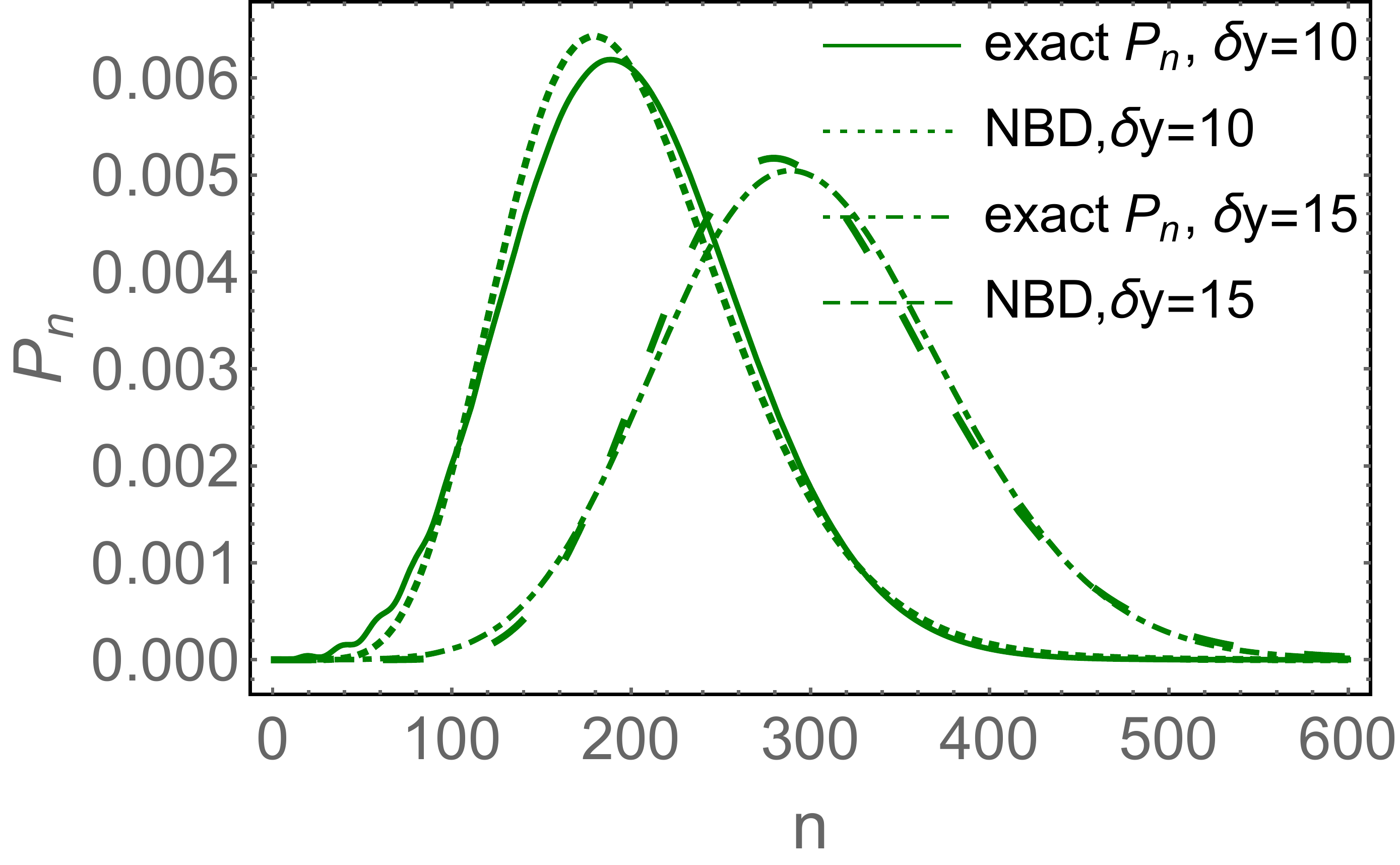}  \\
    \fig{md}-c & \fig{md}-d\\        
  \end{tabular}
        \caption{Multiplicity distributions for two values of  $Y$. The exact $P_n$  denote the distribution of \eq{NH10}, the approximate $P_n$ are the multiplicity distributions in which we sum  over $k$ from $k=0$ up to $k_{max}=2 k_{SP}$.  \
        PD is the Poisson distribution of \eq{PD}. ND and NBD (see text) denote the normal and negative binomial distributions. $  \frac{\Delta}{\gamma}  $ is taken to be equal to 20. $\delta y$ denotes the amount of evolution from the initial rapidity $Y_0$ to $Y$, see \eq{SOL0}.}
\label{md}
   \end{figure}

  It is tempting to estimate the sum in \eq{NH10} by the method of steepest descent, however 
  it turns out that the maximum in $k$ is very broad and the steepest descent method is not applicable. Nevertheless \eq{NH11} gives a good estimate of the important range of $k$.  Limiting summation over $k$ in \eq{NH10} by $k_{max}$, we find that  $k_{max} = 2 k_{SP}$ gives a very good agreement with the exact sum (see \fig{md}).  
  


We plot the distribution given by \eq{NH10} in \fig{md} and compare it to the Poisson distribution \eq{PD} with the same mean value. This latter distribution as discussed previously, is the asymtotic distribution in UTM. Clearly  the distribution of \eq{NH10}  is much broader. On the other hand \fig{md}-c illustrates that \eq{NH10} is well approximated by the normal distribution (ND) with the mean value $<n>=\frac{\Delta}{\gamma}Y$ and  variance $\sigma^2\,=\,\Lb \frac{\Delta^2}{\gamma^2}\,\,+\,\,\frac{\Delta}{\gamma}\Rb\,Y$ as suggested by \eq{NH7},
\beq
P^{ND}_n \,\,=\,\,\frac{1}{\sqrt{2\, \pi\,\sigma^2}}\, \exp\Lb - \frac{ \Lb n \,-\,<n>\Rb^2}{ 2\,\sigma^2}\Rb\eeq
 \fig{md}-d  demonstrates  that an equally good approximation is provided by the negative binomial distribution
  \beq \label{NBD}
 P^{NBD}_n \,=\,\frac{\left(\frac{r}{N+r}\right)^r \left(\frac{N}{N+r}\right)^n \Gamma (n+r)}{n! \Gamma (r)} 
 \eeq 
   with the mean value  $N =M_1^{\mbox{\tiny UTMM}}$ and  parameter $r = \frac{M^2_1}{M_2 - M^2_1}\,\,=\,\,Y$. In fact $P^{NBD}$ reproduces the first and the second terms  in  all $M_k^{\mbox{\tiny UTMM}} $ of \eq{NH7}., viz: $M_k^{\mbox{\tiny UTMM}}\,=\,N^k \Lb 1 + \frac{k (k-1)}{2} \frac{1}{Y}\,\,+\,\,O\Lb \frac{1}{Y^2}\Rb \Rb$.  
 
 As expected therefore, allowing multiple dipole emissions results in a qualitatively different, and much broader distribution. This is true both at intermediate rapidities, where the saturation is unimportant, and also  at asymptotically large energies where the saturation effects play crucial role. Interestingly, in both regimes the average number of dipoles at a given rapidity is the same in UTM and UTMM, and it is the shape of the distribution that discriminates decisively between the two models.

  \section{"Nuclei" and evolution: the $m$-dipole initial condition}
 
 In this paper our primary interest is in the regime where multiple emissions in the evolution are important.  For an initial condition of a single dipole this regime is achieved only at large rapidities. It is interesting to consider how the situation changes if our initial condition itself contains multiple dipoles. Such an initial condition is a proxy to a "nucleus" in the toy world. We expect naturally, that in this case the asymptotic regime in the evolution will be achieved at much lower rapidities. In this section we repeat the analysis of the evolution for the initial condition of exactly $m$ dipoles at initial rapidity. In terms of the initial probability distribution this means
 \beq
 P_{n(m)}(Y=0)=\delta_{nm}
 \eeq
 while in terms of factorial moments:
 \beq
 M_{k(m)}(Y=0)=\frac{m!}{(m-k)!}; \ \ \ k\le m; \ \ \ \ \ M_{k(m)}(Y=0)=0;\ \ k>m
 \eeq
 or in terms of the wave function
 \beq\label{zin}
 Z_0(u)=u^m
 \eeq
 \subsection{Many dipoles evolved with BK}
For the BK evolution the solution for the $m$-dipole initial condition is easy to find. Using \eq{zsol} and \eq{zin} we find 
\beq
Z^{BK}_Y(u;m)=\left[\frac{u}{u(1-e^{\Delta Y})+e^{\Delta Y}}\right]^{m}.
\eeq
With this generating function the probabilities are found to be
\beq  \label{PNM}
P_{n(m)}^{BK}(Y)=C^{n-1}_{m-1}e^{-m\Delta Y}\left[1-e^{-\Delta Y}\right]^{n-m}.
\eeq
The binomial coefficient $C^{n-1}_{m-1}$ here is simply the number of ways to put $n$ identical objects into $m$ boxes without leaving a single box empty.

The first moment is also easily calculated
\beq\label{MBM1}
M^{BK}_{1(m)}\equiv N_{(m)}(Y)=me^{\Delta Y}
\eeq
while for the $k$-th moment we get
\beq
M^{BK}_{k(m)}=k!e^{k\Delta Y}\sum_{l=1}^{m}C_l^mC_{l-1}^{k-1}\left[1-e^{-\Delta Y}\right]^{k-l}\,\,=\,\,k!\,\, e^{ \Delta\,k\,Y}\,\,m \,T^{k-1} \, _2F_1\left(1-k,1-m;2;\frac{1}{T}\right)
\eeq
where $T\,\,=\,\,1\,\,-\,\,\exp\Lb - \Delta\,Y\Rb\,\,=\,\,1\,-\,\frac{m}{N_{(m)}}$.
For large rapidities
   \begin{subequations}  
\bea\label{largeY}
P_{n(m)}^{BK}(Y)&\xrightarrow{\Delta\,Y\,\gg\,1}&\frac{(n-1)!}{(n - m)!\,(m-1)!}\left(\frac{m}{N_{(m)}(Y)}\right)^m \,\,\exp\left[-\frac{m(n-m)}{N_{(m)}(Y)}\right]\nonumber\\
&\xrightarrow{n\gg m; \ e^{\Delta Y}\gg m}&\frac{m}{(m-1)!}\frac{1}{N_{(m)}(Y)}\,
\left(\frac{mn}{N_{(m)}(Y)}\right)^{m-1}\exp\left[-\frac{mn}{N_{(m)}(Y)}\right]\label{PKBKM}\\
M^{BK}_{k(m)}&\!\!\!\xrightarrow{ e^{\Delta \,Y}  \gg m}& \,\,\left(\frac{N_{(m)}(Y)}{m}\right)^k\,\frac{ (m+k-1)! }{(m-1)!  \label{MKBKM}}
\eea
 \end{subequations} 

We note that for $m=1$ the probability distribution at high energy satisfies the KNO scaling. For $m>1$ this is strictly speaking not true, however at high enough energy where $\exp\{\Delta Y\}\gg m$ and only large values of $n\gg m$ matter for the bulk properties of the distribution the KNO scaling is restored as is clear from eq.(\ref{largeY}). Nevertheless even at high energies the initial number of dipoles $m$ appears as a parameter in the KNO function. In particular the KNO function drops faster at large values of the argument for large $m$.

\subsection{$m$-dipoles in UTM at intermediate rapidities}

The initial condition determines matching of the solution of UTM with the BK regime. 
 Matching \eq{PL8} with eq.(\ref{PKBKM}) at small $Y$ we obtain
  \bea
   P^{\mbox{\tiny  UTM}}_{n(m)}\Lb Y\Rb\,\,&=&\,\,\frac{1}{(m-1)!}\frac{\gamma}{\Lb 1 - e^{- (n - 1) \gamma}\Rb}\,\exp\Bigg( - e^{\zeta(Y,n)} \,\,+\,\,m\zeta(Y,n)\Bigg)\nn\\
  &=&
   \frac{1}{(m-1)!}   e^{ -\,\Delta\,Y\,\,+ \gamma (n -1)} \exp\Bigg( - e^{\zeta(Y,n)} \,\,+\,\,(m\,-\,1)\zeta(Y,n)\Bigg)   ; \label{PL1001}
   \eea
 This qualitatively is rather similar to the solution for $m=1$. In fact the sensitivity to the value of $m$ is weaker in this regime than for the BFKL cascade. 
 
 The value of $n$ at which the probability  \eq{PL1001} is maximal is determined with good accuracy from the relation
 \beq \label{NMAX}
\zeta(Y,n) \,\,=\,\,\ln(m)
\eeq
or from vanishing of the derivative
\begin{eqnarray}\label{p2pm}
\frac{\partial P^{\mbox{\tiny  UTM}}_{n(m)}}{\partial n}&=&\left[\gamma\left[1+\frac{m-1}{1-e^{-(n-1)\gamma}}\right]-e^{-\Delta Y+(n-1)\gamma}\right]P^{\mbox{\tiny  UTM}}_{n(m)};
\end{eqnarray}
Taking $m$ as a number which is parametrically not large $m\ll 1/\gamma$, we find that at large 
  rapidity the maximum of the probability distribution is at
   \beq\label{nmmax}
n^{(m)}_{max}-1\approx\frac{\Delta}{\gamma}Y-\frac{1}{\gamma}\ln\frac{1}{m\gamma}
\eeq
This relation clearly exhibits effects of saturation: the value of $n^{(m)}_{max}$ grows only logarithmically with $m$, whereas if we were to continue the BFKL cascade to these values of rapidity, the position of the maximum would be linear in $m$.

For the mean multiplicity we find 
 \bea \label{PL12}
&&N_{(m)}(Y)\equiv M^{\mbox{\tiny  UTM}}_{1(m)}(Y)\approx\\\
&&\frac{1}{(m-1)!}\Lb - \frac{d\,}{d\,\alpha}\Rb^{m - 1}\left[\frac{e^{\frac{\alpha}{\gamma} e^{-\Delta Y}}}{\alpha\,\gamma}\Bigg( \Gamma\Lb 0, \frac{\alpha}{\gamma}e^{-\Delta Y}\Rb + (m-1)\,\gamma\,e^{ - \frac{\alpha}{\gamma}e^{\Delta\,Y}\exp\Lb( m-1)\,\gamma\Rb}\Bigg)
\right]\Bigg{|}_{\alpha=1}\nn
 \eea 
 
 At small $Y$, the multiplicity $N_{(m)}(Y)$ from \eq{PL12} tends to the BFKL cascade value $ m \,e^{\Delta\,Y} $. At large $Y$ \eq{PL12} asymptotically gives $N_{(m)}(Y)\,\,\to\,\,\frac{\Delta}{\gamma} \,Y\,\,-\,\, \frac{1}{\gamma}\ln\frac{1}{\gamma} $. To reproduce the weak logarithmic dependence on $m$ of Eq.({\ref{nmmax} ) one would need to keep sub asymptotic terms in the expansion of the incomplete gamma function. At very large $Y$ these corrections are unimportant and we recover the same $m$-independent asymptotics as given by  $Z_Y^{asymp}\Lb u\Rb$ Eq.(\ref{SOL0}).

\subsection{UTM at asymptotically large rapidities}
If $m$ is not large parametrically, the exact value of $m$ is not very important for the probability distribution in the saturation regime. The onset of the saturation regime is a little earlier than for $m=1$, as follows from \eq{nmmax}, but the distribution itself is practically the same. However if $m$ is very large, i.e. $m\sim1/\gamma$ the saturation corrections in the evolution kick in right away.  In this case there is no BK or intermediate regime in the evolution and already at low rapidity one can approximate the evolution by \eq{schsat}. The solution at arbitrary rapidity then has the form
\beq \label{SOL1}
Z_{Y(m)}^{UTM}\Lb u\Rb\,\,=\,\,e^{  \frac{\Delta}{\gamma}( u - 1) \,Y}u^m
\eeq
 The ensuing dipole distribution is Poisson shifted by the initial number of dipoles
 \beq \label{PDm}
P^{(m)}_n\Lb Y\Rb\,\,=\,\,\frac{(\frac{\Delta}{\gamma}Y)^{(n-m)}}{(n-m)!}e^{-\frac{\Delta}{\gamma}Y}
\eeq
The mean multiplicity in this cascade
\beq N_{(m)} = \frac{\Delta}{\gamma}\,Y+m\eeq

The influence of the initial number of dipoles $m$  is illustrated by \fig{nmax} where  we plot the value of $N_{(m)}$.  
In this figure the two limiting cases are also shown: the value of $N_{(m)}$ for the BFKL cascade  $N_{(m)}=m\exp\Lb \Delta\,Y\Rb$, and \eq{nmmax}, which 
gives the maximum probability in the saturation region.  It is clear from this figure that the transitional region of $Y$ between the BFKL cascade and the asymptotic multiplicity distribution in the saturation region for the UTM cascade  shrinks  for large $m$. The values of $m$ that are shown in \fig{nmax} , were chosen  having in mind $m=3$ for the "proton", and $m =3A^{1/3}$ for a "nucleus" of atomic number $A$.

    \begin{figure}[ht]
    \centering
  \leavevmode
      \includegraphics[width=15cm]{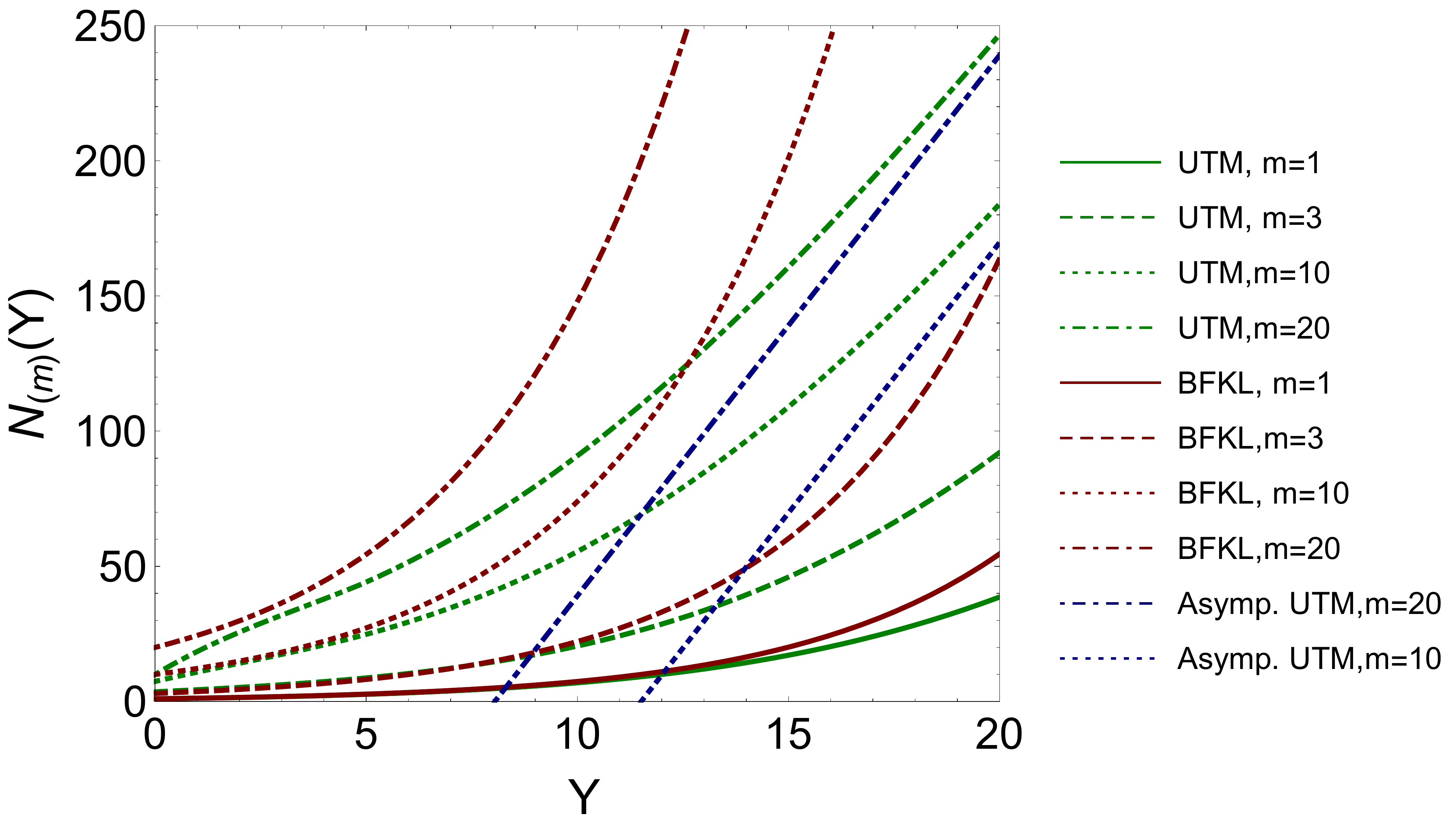}  
   \caption{  The value of $N_{(m)}$ from \protect\eq{PL12} versus $Y$.  In red it is shown the value  for the BFKL cascade (see \eq{MBM1}) 
   while in blue the values of $n_{max}$  from \eq{nmmax} are indicated. }
\label{nmax}
  \end{figure}


 \subsection{UTMM - multiple emissions without saturation}
 
We can now repeat the analysis for UTMM.  \eq{MGF4} and \eq{MGFPN} give the factorial moments and $P_n$ for single dipole initial condition.
 For the $m$-dipole case,  \eq{smalln} gives $P^{\mbox{\tiny UTMM}}_{n<m}=0$ . $P^{\mbox{\tiny UTMM}}_{m} $ and $M^{\mbox{\tiny UTMM}}_1$  
 are easily determined from this equation,  and from \eq{M1}: 
 \beq  \label{MGF10}  
 P^{\mbox{\tiny UTMM}}_{n \,<\,m}\Lb Y \Rb  \,\,=\,\,0;~~~~~~ P^{\mbox{\tiny UTMM}}_{n \,=\,m}\Lb Y \Rb  \,\,=\,\,
 e^{-\,\Delta\,Y};~~~~~M^{\mbox{\tiny UTMM}}_{1\,(m)}\Lb Y\Rb\,\,\,=\,\,m\,\,e^{\,\Delta\,Y}.
 \eeq
 The next two moments are determined from \eq{M2} and \eq{M3} as:
 \bea \label{MGF11}
 M^{\mbox{\tiny UTMM}}_{2\,(m)}\Lb Y\Rb\,\,&=&\,\,m \frac{ (2 + \Delta)}{  (\Delta + 1)}\Bigg( e^{( 2  + \Delta)\Delta\,Y} \,\,-\,\,e^{\Delta\,Y}\Bigg)\,\,+\,\,m \,\Lb m - 1\Rb \, e^{( 2  + \Delta)\Delta\,Y}\nn\\
 &\xrightarrow{\Delta \,\ll\,1}& 2\,m \Lb e^{\Delta_2\,Y}\,\,-\,\,e^{\Delta_1\,Y} \Rb\,\,+\,\,m \,(m \,-\,1)\,e^{\Delta_2\,Y}
  \eea  
 
 \bea \label{MGF12}
 M^{\mbox{\tiny UTMM}}_{3\,(m)}\Lb Y\Rb\,\,&=&\,6\,e^{ \Delta_3\,Y} \int^Y_0 d Y' \,e^{- \,\Delta_3\,Y'}\, M_2\Lb Y'\Rb\,\,\,+\,\,m\,(m \,- \,1)\,(m\,-\,2 )\,e^{ \Delta_3\,Y} \nn\\
 & = &3!\,m \Bigg( e^{\Delta_3\,Y} \,\,-\,\,2 \,e^{\Delta_2\,Y}\,\,+\,\,e^{ \Delta\,Y}\Bigg)\,\,+\,3!\,m \,(m \,-\,1)\,\Bigg(e^{\Delta_3\,Y} \,-\, e^{\Delta_2\,Y}\Bigg)\nn\\
 &\,\,+&\,\,m\,(m \,- \,1)\,(m\,-\,2 )\,e^{ \Delta_3\,Y} 
 \eea  
 where, as before $\Delta_k \,\,=\,\,(1\,+\,\Delta)^k \,\,-\,\,1$.
For  $M_{4\,(m)}^{\mbox{\tiny UTMM}}$  we get:
 \bea \label{MGF13}
 M^{\mbox{\tiny UTMM}}_{4\,(m)}\Lb Y\Rb\,\,&=&\,12\,e^{ \Delta_4\,Y} \int^Y_0 d Y' \,e^{- \,\Delta_4\,Y'}\, M_3\Lb Y'\Rb\,\,\,+\,\,m\,(m \,- \,1)\,(m\,-\,2 )\,(m - 3)\,e^{ \Delta_4\,Y} \nn\\
 & = &4!\,m \Bigg( e^{\Delta_4\,Y} \,-\,3 \,e^{\Delta_3\,Y}\,+\,3\,e^{ \Delta_2,Y}\,-\,e^{\Delta_1\,Y}\Bigg)\,+\,m \,(m \,-\,1) \,4! \frac{3}{2}\,\Bigg(e^{\Delta_4\,Y} \,-\, 2\,e^{\Delta_3\,Y}\,+\,e^{\Delta_2\,Y}\Bigg)\nn\\
 &\,\,+&\,\, 4!\, \h  \, m\,(m \,- \,1)\,(m\,-\,2 )\,\Bigg(e^{ \Delta_4\,Y}  \,-\, e^{ \Delta_3\,Y}\Bigg)\,\,+\,\,        
 m\,(m \,- \,1)\,(m\,-\,2 )\,(m - 3)\,e^{ \Delta_4\,Y} \eea 
 These expressions look rather complicated. However at large rapidities the coefficient of the leading exponent in the $k$-th moment is easy to determine.
 We note that at small $Y$ the solution should reduce to that of the BFKL cascade, since multiple emissions are unimportant for small $Y$.  Since the structure of the moments in both regimes is a linear combination of exponentials, when setting $\Delta_k \,=\, k \,\Delta$, the moments for UTMM without saturation should reduce to the BFKL moments. This allows us to calculate the combinatorial factor in $M^{\mbox{\tiny UTMM}}_{k(m)}$ by using the result of the BFKL cascade \eq{MKBKM}:
 \beq
 M^{\mbox{\tiny UTMM}}_{k(m)}\approx \frac{(m\,+\,k\,-\,1)!}{(m\,-\,1)!}e^{\Delta_kY}.
 \eeq
 One can check directly that this reproduces the coefficient of the highest exponent in \eq{MGF11}, \eq{MGF12} and \eq{MGF13}.
Using this expression for the moments we can write the generating function as
  \beq \label{MGF16}
 Z^{\mbox{\tiny UTMM}}_Y\Lb u;m \Rb\,\,=\,\,1\,\,+\,\,\sum_{k=1}^\infty \frac{(m\,+\,k\,-\,1)!}{(m\,-\,1)!\,k!}\,(u-1)^k \,e^{\Delta_k\,Y}.
 \eeq 
  We now expand this equation with respect to $(1 + \Delta)^k\,Y$:
  \beq  \label{MGF17} 
 Z^{\mbox{\tiny UTMM}}_Y(u;m) \,=\,e^{-Y} \,\sum^{\infty}_{j=0} \sum^{\infty}_{k=0}\,\frac{(m+k-1)!}{(m-1)!\,k!}\, \frac{(1+\Delta)^{k\,j}\,Y^j }{j!}\,(u-1)^k\,=\,e^{-Y} \sum^{\infty}_{j=0} \frac{Y^j }{j!}\,\Bigg(\frac{1}{ 1 - \Lb 1+\Delta\Rb^j \,(u-1)}\Bigg)^{m}
  \eeq 
 Reexpanding this  in powers of $u^n$, we obtain 
  the following representation for the probability distribution $P_n^{\mbox{\tiny UTMM}}(Y)$:
  \beq \label{MGF18}
P^{\mbox{\tiny UTMM}}_n\Lb Y\Rb  =\,\,e^{-Y}\sum^{\infty}_{j=0} \frac{Y^j}{j!} \,P^j_n~~~~~\mbox{with}~~~
P^j_n\,\,=\,\ \frac{(m\,+\,n\,-\,1)!}{(m\,-\,1)!\,n!}\Bigg(\frac{1}{\Lb N_j+1\Rb^{m} } \Lb 1 + \frac{1}{N_j}\Rb^{-n }\Bigg) \eeq  
 where $N_j\,\,=\,\,\Lb1\,+\,\Delta\Rb^j$.
 
 The multiplicity distribution for \eq{MGF18} is plotted in \fig{pnn0} at different values of $m$ for $Y =10$.


     \begin{figure}[ht]
    \centering
  \leavevmode
  \begin{tabular}{c c}
      \includegraphics[width=8.1cm]{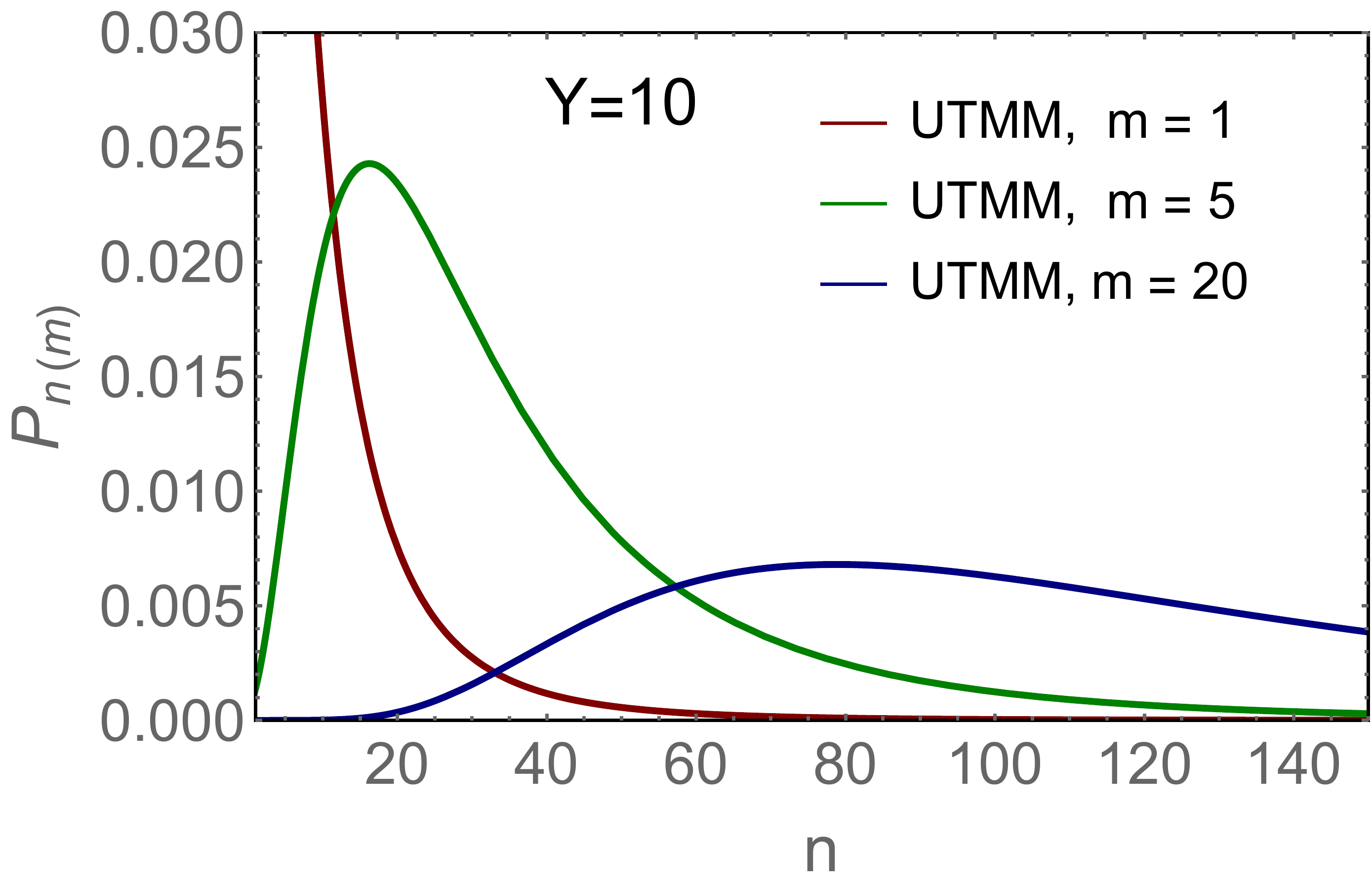}  &  \includegraphics[width=8cm]{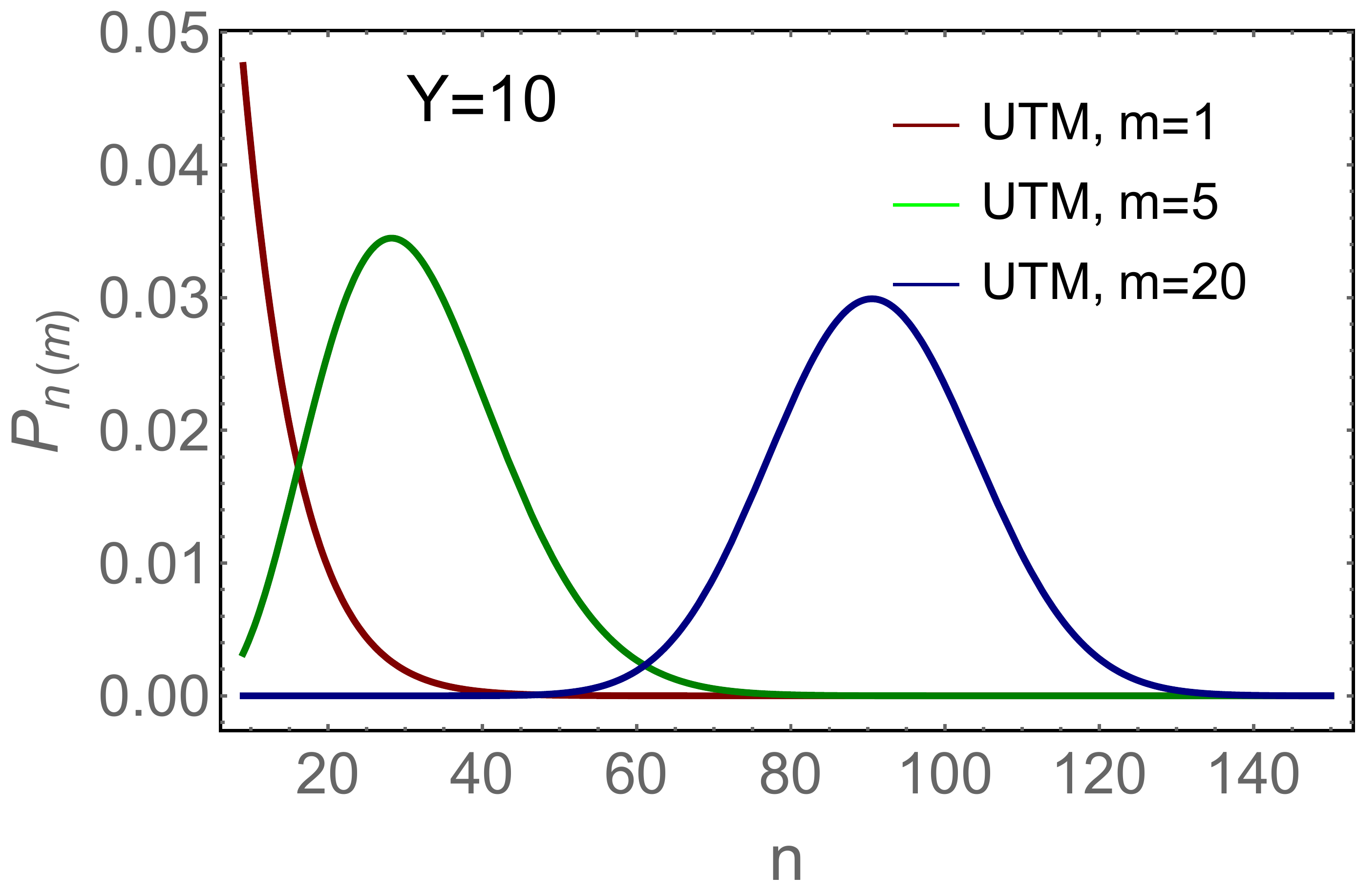}\\
      \fig{pnn0}-a & \fig{pnn0}-b\\
      \end{tabular}     
       \caption{\fig{pnn0}-a:  $P_n$  from \eq{MGF18} (UTMM cascade)   versus $n$ at different values of $m$.  \fig{pnn0}-b:  $P_n$  for the UTM cascade. $\Delta = 0.2,\,\,
        \gamma = 0.01$.}
\label{pnn0}
   \end{figure}

\subsection{UTMM with saturation}
Finally, when $m$ is large the large $Y$ of UTMM is modified similarly to UTM. In particular the generating function of \eq{NH5} is modified as
\beq
  Z^{\mbox{\tiny UTMM}}_{asymp}\,\,=\,\,\exp \Lb\Bigg( e^{ \frac{\Delta}{\gamma} \Lb u  -1\Rb} \,\,-\,\,1\Bigg)\,\,Y\Rb u^m
\eeq
 and the probability distribution is shifted by $m$
 \beq
 P^{UTMM}_{n(m)}(Y)=P^{UTMM}_{n-m(1)}
(Y).
\eeq 
 
 \section{Discussion}
 This paper has one central point. For large number of partons in hadronic wave function, rapidity evolution should not be limited by emission of a single parton in one step. Instead many partons can be independently emitted. Here in the framework of a toy model in zero transverse dimensions we have implemented this idea by constructing an evolution which describes independent emission of multiple partons (dipoles).  This evolution by construction preserves $t$-channel and $s$-channel unitarity. It incorporates saturation dynamics, and reduces to the known simpler models whenever multiple emissions are unimportant.
 The model has two parameters: the dipole emission probability $\Delta$, and dipole-dipole elastic scattering amplitude $\gamma$. Within the model itself these parameters are independent, but when projected on QCD we expect $\Delta\sim \alpha_s$ and $\gamma\sim \alpha_s^2$.
 
 In this sense this is an appropriate evolution to be applied for scattering of dense objects, unlike the models of similar type considered so far. 
 We have studied various aspects of probability distributions generated by this evolution from two type of initial conditions: a single dipole and multiple dipoles. An interesting feature of the model worth noting is that the regime where multiple scatterings are important precedes the onset of saturation corrections. Parametrically, as long as the average number of dipole is small $N<1/\Delta\sim 1/\alpha_s$, the single dipole emission dominates the evolution and the probability distribution is that of the BFKL cascade. For intermediate values of $N$ such that $1/\Delta<N<1/\gamma$ saturation effects are still unimportant, but multiple dipole emissions are dominant. Finally at very large rapidities where $N>1/\gamma$ both, multiple emissions and saturation corrections determine the asymptotic dipole distribution.
 
 There are significant differences between the asymptotic behavior of multiplicity distributions in the model that allows only emission of a single dipole (UTM) and that with multiple dipole emission (UTMM). The main qualitative difference as illustrated on \fig{md}, is that the distribution in UTMM is significantly wider, and is well approximated by the normal distribution whereas in UTM the asymptotic distribution is of the Poisson type.
 One can understand this feature directly from the asymptotic form of the generating function $Z$. Interestingly, in both models the asymptotic distribution can be written in the form
 \beq\label{scale}
 Z_Y(u)=(z(u))^Y
 \eeq
 with
 \beq\label{fundc}
 z_{UTM}(u)=e^{ \frac{\Delta}{\gamma} \Lb u  -1\Rb }\,\, \ \ \ \ \ \ \ \ \ \ z_{UTMM}(u)=e^{\left[e^{ \frac{\Delta}{\gamma} \Lb u  -1\Rb} \,\,-\,\,
 1\right]}=\sum_{k=0}^\infty \frac{1}{k!}(z_{UTM}-1)^k
 \eeq
 Our discussion in the previous section makes it clear that taking a power $m$ of a distribution function is equivalent to considering a state that evolves into $m$ independent cascades, so that the final probability distribution is that of $m$ cascades. From this point of view the rapidity $Y$ in \eq{scale} plays the role of the number of independent cascades in the asymptotic distribution. 
 
 It is therefore natural to interpret the asymptotic probability distribution in the following way. At pre asymptotic rapidities $Y<Y_{asymp}$ an initial state evolves into some fundamental distribution, or cascade $z$. Starting from $Y_{asymp}$ the asymptotic evolution takes over, which amounts simply to multiplication of the number 
 of these independent fundamental cascades at a constant rate. At any rapidity $Y\gg Y_{asymp}$ the number of such fundamental cascades is $m\approx Y$. Different evolution dynamics correspond to different properties of the fundamental cascade $z$ \eq{fundc}. In UTM $z_{UTM}$ is a Poisson distribution with average dipole number $\langle n\rangle =\frac{\Delta}{\gamma}$.  A composition of $Y$ independent Poisson distributions gives again a Poisson distribution with the additive mean value $N(Y)=\langle n\rangle Y$, which is precisely what we have seen in Section 2. On the other hand in UTMM, since the pre asymptotic evolution is dominated by multiple dipole emission, the fundamental distribution $z_{UTMM}$ is not a Poisson, but rather a weighted sum of Poisson distributions with averages which are multiples of $\langle n\rangle$. A large number of such cascades is not a Poisson distribution anymore. On the other hand we expect that a composition of a large number of identical distributions must lead to a normal distribution, which explains why the distribution on \fig{md} 
is so close to a normal distribution.
 
 In general we saw that including the saturation effects changes the shape of the distribution significantly. In particular while the BK distribution at large $Y$ satisfies KNO scaling, the asymptotic UTM distribution does not. The KNO property is interesting in relation to the recent discussions of parton entropy pioneered in 
 Ref.\cite{KHLE} .  It was noted in \cite{KHLE}  that at large average multiplicities $N$ the entropy of the BFKL cascade 
 behaves as 
 \beq \label{C1} 
  S=\ln N.
  \eeq
  Such behavior can be interpreted in terms of a large number of partonic micro-states having equal probabilities. The proton then is thought of as composed of  an exponentially large (in rapidity) number $N$ of micro-states that occur with equal and small probabilities $1/N$. More generally one can see that any probability distribution that follows KNO scaling  $P_n = \frac{1}{N}\Psi\Lb \frac{n}{N}\Rb$, yields logarithmic entropy
  \beq\label{ent}
  S=-\sum_nP_n\ln P_n\approx -\int \frac{dn}{N}\Psi(\frac{n}{N})\left[-\ln N+\ln \Psi(\frac{n}{N})\right]=\ln (aN)
  \eeq
  where $a$ is a rapidity independent constant. Strictly speaking, of course not any KNO-type distribution corresponds to equally populated micro states, but physically the situation is not too different. As $N$ grows, KNO scaling means that more and more states (wider range of $n$'s) get populated, if not exactly equally at least according to some fixed probability  ratio determined by the KNO function $\Psi$.
 This is the origin of the logarithmic term in the entropy.
 
   In this paper we demonstrated that both  the cascade  with saturation, and a possibility to emit many partons lead to violations of  KNO scaling. It would be interesting to see to what extent this also leads to violation of  \eq{C1}. We can answer this question for asymptotic rapidities. Both in UTM and UTMM at very large rapidities the distribution is well approximated by normal distribution with $\sigma^2\propto N$. Estimate similar to \eq{ent} then gives
   \beq
   S=\ln a\sigma=\frac{1}{2}\ln N+const
   \eeq
 This differs from \eq{C1} by a factor of two, which reflects the difference scaling property of the distribution with $N$.

   The second point we want to stress is that at intermediate rapidities the distribution with saturation is poorly described by the negative binomial distribution of \eq{NBD}, see \fig{comp2}. Within the CGC approach the multiplicity distribution was calculated in the MV model, the so called 
   "glittering glasma"\cite{GLPL}. This leads to  NBD in which the average number of partons and parameter $r$ are determined by the first and the second factorial moments. Both UTM and UTMM are far from NBD at intermediate rapidities, although interestingly in the asymptotic regime the UTMM cascade is quantitatively not very different from NBD \fig{md}-d.   
     \begin{figure}[ht]
    \centering
  \leavevmode
  \begin{tabular}{c c}
      \includegraphics[width=8.1cm]{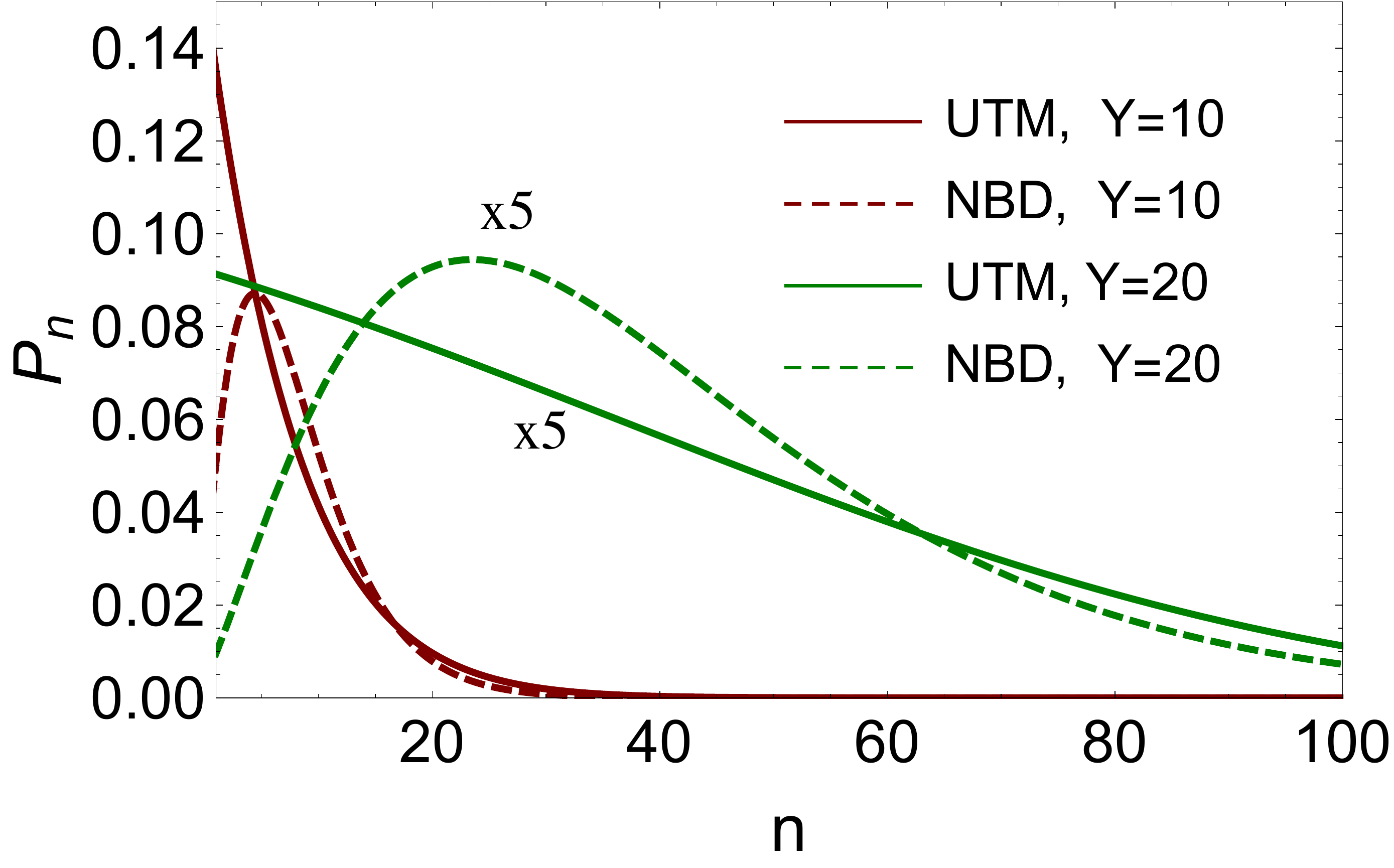}  &  \includegraphics[width=8cm]{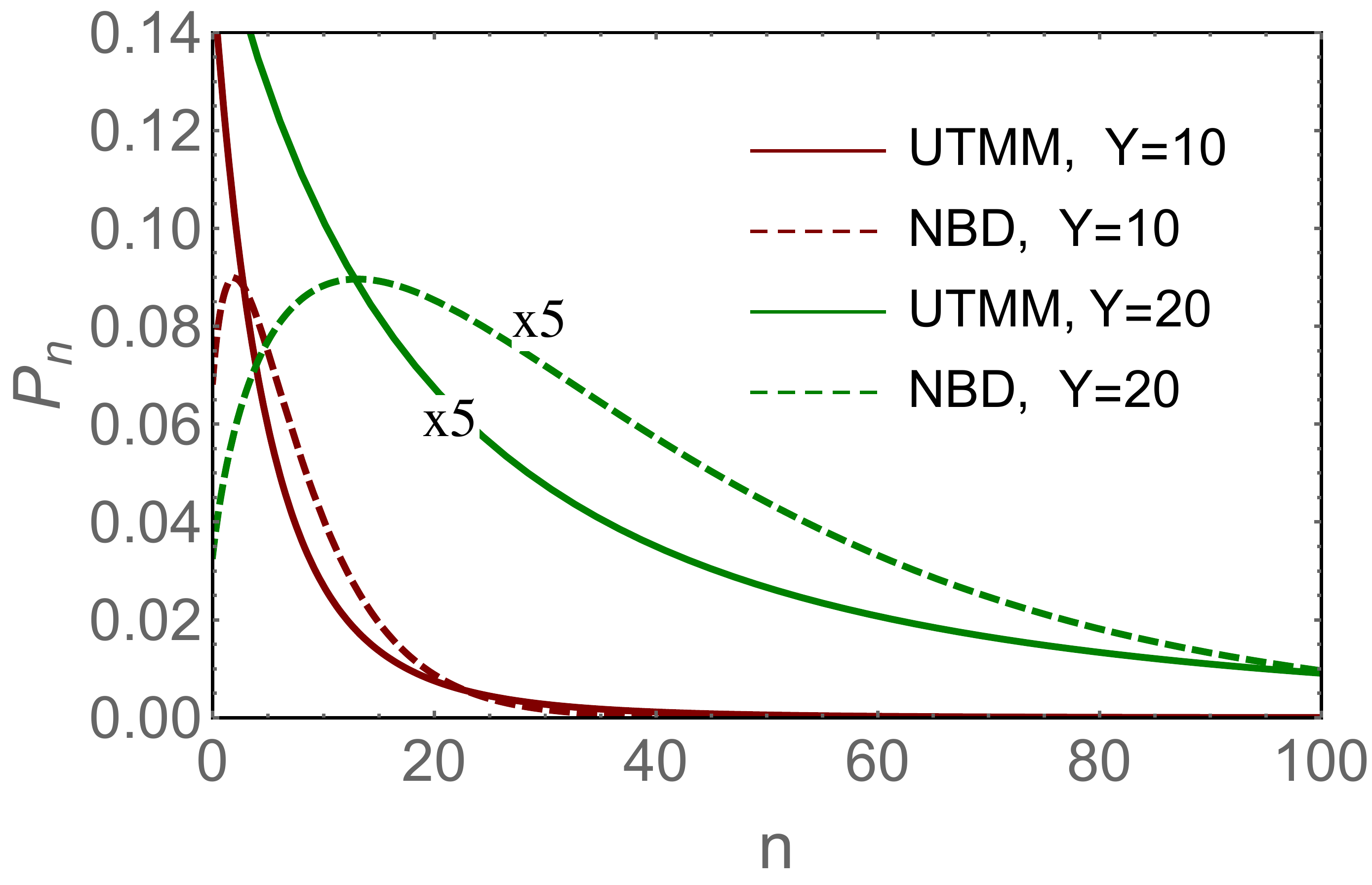}\\
      \fig{comp2}-a & \fig{comp2}-b\\
      \end{tabular}     
       \caption{Comparison between multiplicity distributions in UTM and UTMM cascade with the negative binomial distributions which have the same $M_1$ and $M_2$ as the cascades. At the plots, $P_n$ are multiplied by 5 for Y=20, to fit the scale. $\Delta =0.2,\gamma = 0.01$. }
\label{comp2}
   \end{figure}

 We believe the main point of this paper is valid beyond the toy model and generalizes to high energy scattering in QCD as well. The BK regime in QCD should not transition directly into the saturation regime as far as the evolution of the wave function is concerned. Rather the saturation should be preceded by the rapidity range where multiple gluon emissions in the evolution play prominent role. The study of this new regime is an interesting problem which should start by the derivation of the generalization of the BK Hamiltonian.
 

     \appendix
 \section{Factorial moments and probabilities in UTMM}
 
 Here we derive the approximate result for the factorial moments of the distribution which generalizes eq.(\ref{M32}).
    
 Using $M_k^{\mbox{\tiny UTMM}} \,= \frac{d^k \,}{d\, u^k} Z^{\mbox{\tiny UTMM}}_Y(u)|_{u=1}$ and the Schroedinger equation \eq{SE1} we obtain the equation for the factorial moments
  \bea \label{MGF1} 
\frac{d \,M^{\mbox{\tiny UTMM}}_k(Y)}{d\,Y} & \,=\,&\frac{d}{d\,Y}  \frac{d^k}{d u^k}Z^{\mbox{\tiny UTMM}}_Y(u)|_{u=1} \,\,=\,\,
\Bigg(\frac{d^k}{d u^k} \sum^\infty_{n=1}\,\Delta^n\,(u -  1)^n \Lb u \frac{d}{d u}\Rb^n Z^{\mbox{\tiny UTMM}}_Y(u)\Bigg)\Bigg{|}_{u=1}  \nn\\
 &=&   \Bigg(\frac{d^k}{d u^k} \sum^{k}_{n=1}\,\Delta^n\,(u - 1)^n \Lb u 
 \frac{d}{d u}\Rb^n Z^{\mbox{\tiny UTMM}}_Y(u)\Bigg)\Bigg{|}_{u=1} 
 \eea
 With some arduous algebra the right hand side of this equation can be rewritten in terms of the moments $M_m^{\mbox{\tiny UTMM}}$. This expression is a linear function of all $M_j^{\mbox{\tiny UTMM}}$ with  $j\le k$. We know however that at small $\Delta$ and large rapidity the higher moments are exponentially larger than the lower ones. Thus at high rapidity we do not need to keep all the terms in the RHS. The largest term is the one that is proportional to $M_k^{\mbox{\tiny UTMM}}$. Keeping only this term would give us a homogeneous equation for $M_k^{\mbox{\tiny UTMM}}$, and although we would be able to determine the largest term at high $Y$, we would not be able to impose the correct initial condition. We therefore also keep the next largest term which contributes to the non homogeneous term  in \eq{MGF1}. We will however disregard perturbative in $\Delta$ corrections to the prefactors in the exponentials. This makes it possible to simplify \eq{MGF1} by substituting $u\frac{d}{du}\rightarrow \frac{d}{du}$ in all but $n=1$ terms
 \bea \label{MGF2} 
 \frac{d \,M^{\mbox{\tiny UTMM}}_k(Y)}{d\,Y}\,\,&\approx&\,\,  \Bigg(\frac{d^k}{d u^k}\Lb \Delta\,(u-1) u\frac{d}{d u} +  \sum^{k}_{n=2}\,\Delta^n\,(u - 1)^n \Lb 
 \frac{d}{d u}\Rb^n\Bigg) Z^{\mbox{\tiny UTMM}}_Y(t)\Rb\Bigg{|}_{u=1} \nn\\
  &=& \Delta_k \,M^{\mbox{\tiny UTMM}}_k\Lb Y\Rb \,+\,\Delta\,k(k-1)\,M^{\mbox{\tiny UTMM}}_{k-1}\Lb Y\Rb
  \eea
 where $\Delta_k\,\,=\,\,( 1 + \Delta)^k\,-\,1$. The solution to \eq{MGF2} with a single dipole initial condition is:
  \beq \label{MGF3}
 M^{\mbox{\tiny UTMM}}_k(Y)\,\,=\,\, \Delta\,k (k - 1)\, e^{\Delta_k\,Y} \int\limits^Y_0 d Y'  M^{\mbox{\tiny UTMM}}_{k-1}\Lb Y'\Rb \,e^{ - \Delta_k\,Y'}
 \eeq
 Note, that \eq{MGF3} determines $M^{\mbox{\tiny UTMM}}_k$ for $k \geq 2$ while for $M^{\mbox{\tiny UTMM}}_1$ we have the solution of \eq{M11}.

Inspired by  \eq{M21} and \eq{M32} we make the induction hypothesis, that $M^{\mbox{\tiny UTMM}}_j$ for $j\,  \leq\,k-1$   has the form:
  \beq \label{MGF4}
 M^{\mbox{\tiny UTMM}}_j(Y)\,\,=\,\, j!\Bigg( \sum^{j-1}_{l=0} (-1)^{ l}\frac{(j -1)!}{(j -1- l)!\,l!} e^{\Delta_{j-l}\,Y}\Bigg)
 \eeq 
 Substituting this into \eq{MGF3} we now prove \eq{MGF4} for $ M^{\mbox{\tiny UTMM}}_k(Y)$.  Substitution of \eq{MGF4} into \eq{MGF3} yields:
   \begin{subequations}  
    \bea 
 \hspace{-1cm}M^{\mbox{\tiny UTMM}}_k\Lb Y\Rb &=&k! \Bigg( \underbrace{\Lb\sum^{k-2}_{l=0} \frac{(-1)^{l} (k-1)!}{(k - 2 - l)!\,l!} \frac{\Delta}{ \Delta_k - \Delta_{l+1}}\Rb}_{ =\,\,\,1} e^{ \Delta_{k}\,Y} +\sum^{k-2}_{l=0} \frac{(-1)^{k-l} (k-1)!}{(k-2 - l)!\,l!} \frac{\Delta}{ \Delta_k - \Delta_{l+1}} e^{\Delta_{k-l},Y}\Bigg)\label{MGF51}\\
 &\approx&k! \Bigg( \sum^{k-1}_{l=0} (-1)^{l}\frac{(k-1)!}{(k-1 - l)!\,l!} e^{\Delta_{k-l}\,Y}\Bigg) \label{MGF52}
  \eea
    \end{subequations}    
    To obtain the second line we have used 
 $ \Delta_k - \Delta_l \,\approx\,(k\, -\, l)\Delta$ and changed the summation index from $l$ to $l+1$.  Within our approximation we are allowed to use the small $\Delta$ limit in the prefactor of each exponent on the right hand side of \eq{MGF52} while keeping the complete expression for $\Delta_k$ in the exponent. We now see that \eq{MGF52} coincides with \eq{MGF4}. Bearing in mind \eq{M1} - \eq{M32} , we conclude that 
 \eq{MGF4} is valid for all $k$.
 
 We next derive approximate expressions for probabilities at large rapidity $Y$.
   Using \eq{MGF4} we can calculate the generating function $Z^{\mbox{\tiny UTMM}}(u)$, since
 \beq \label{MGF6} 
 Z^{\mbox{\tiny UTMM}}_Y(u) \,\,=\,\,1 \,\,+\,\,\sum^{\infty}_{k=1} \,\frac{M^{\mbox{\tiny UTMM}}_k\Lb Y\Rb}{ k!} \,\Lb u \,-\,1\Rb^k
 \eeq
 We take the simplified expression at large  $Y$: $M^{\mbox{\tiny UTMM}}_k \,\,=\,\,k!\,e^{  \Delta_k \,Y}$ (see \eq{MGF52}). Then 
  \beq  \label{MGF8} 
 Z^{\mbox{\tiny UTMM}}_Y(u) \,\,=\,1\,+\,\sum^{\infty}_{k=1} e^{\Delta_k\,Y} \,\Lb u - 1\Rb^k
 \eeq 
 Since $\Delta_k = (1\,+\,\Delta)^k - 1$, the series of \eq{MGF8} is divergent. To sum such an asymptotic series we need to invent analytical function, which has the same series. We suggest the following procedure:  plugging in \eq{MGF8} $\Delta_k$ we can expand with respect to parameter $(1\,+\,\Delta)^k  \,Y$: viz.:
  \beq  \label{MGF81} 
 Z^{\mbox{\tiny UTMM}}_Y(u) \,\,=\,e^{-Y} \,\sum^{\infty}_{j=0} \sum^{\infty}_{k=0} \frac{(1\,+\,\Delta)^{k\,j}\,Y^j }{j!}\,\Lb u - 1\Rb^k\,\,\,=\,\,e^{-Y} \,\sum^{\infty}_{j=0} \,\frac{Y^j }{j!}\,\frac{1}{ 1 - \Lb 1\,+\,\Delta\Rb^j \,\Lb u - 1\Rb}
  \eeq
 Expending \eq{MGF81} with respect to $u^n$, we obtain 
  the following expression for $P_n^{\mbox{\tiny UTMM}}(Y)$:
 
 \beq \label{MGFPN}
P^{\mbox{\tiny UTMM}}_n\Lb Y\Rb  =\,\,e^{-Y}\sum^{\infty}_{j=0} \frac{Y^j}{j!} \,P^j_n~~~~~\mbox{with}~~~
P^j_n\,\,=\,\,\frac{1}{N_j} \Lb 1 + \frac{1}{N_j}\Rb^{-n }\,\xrightarrow{N_j >1,n>1}\, \frac{1}{N_j} \exp\Lb- \frac{n}{N_j}\Rb
 \eeq  
 where $N_j\,\,=\,\,\Lb1\,+\,\Delta\Rb^j$.

  Since we have made some approximations , it is instructive to check that \eq{MGFPN} leads to the factorial moments $M^{\mbox{\tiny UTMM}}_k \xrightarrow{Y \,\gg\,1} \, <|n^k|>\,=\,k!\,\exp\Lb \Delta_k\,Y\Rb$, reproducing \eq{MGF8}.
  

   {\bf Acknowledgements} 
     
   We thank our colleagues at Tel Aviv university and UTFSM for
 encouraging discussions.     AK  was supported by the NSF Nuclear Theory grant 1913890.
  EL was supported  by 
 ANID PIA/APOYO AFB180002 (Chile),  Fondecyt   grant \#1180118 (Chile) and the Tel Aviv university encouragement grant \#5731. 
ML and AK were supported by the Binational Science Foundation grant \#2018722, and the Horizon 2020 RISE
 "Heavy ion collisions: collectivity and precision in saturation physics"  under grant agreement No. 824093.

\end{document}